\documentclass[12pt]{article}
\usepackage[dvips]{graphicx}
\usepackage{pdproc}
\makeatletter
\def\@cite#1{[#1]}
  \makeatother
\begin{document}

\renewcommand{\thefootnote}{\alph{footnote}}


\begin{flushright}
FERMILAB-CONF-04-311-E   \\
CDF/PUB/CDF/PUBLIC/7276  \\
November 12, 2004
\end{flushright}

\vspace*{2cm}

\title{
Electroweak, Top and Bottom Physics at the Tevatron
}

\author{ FUMIHIKO UKEGAWA  (CDF Collaboration)}

\address{ 
Institute of Physics, University of Tsukuba \\
Tennoudai 1-1-1,  Tsukuba-shi,  Ibaraki-ken 305-8571, Japan
\\ {\rm E-mail: ukegawa@hep.px.tsukuba.ac.jp
 \\ {\mbox {}} \\ representing the CDF and D0 collaborations }}

\abstract{
The Tevatron Run-II program has been in progress since 2001,
and the CDF and D0 experiments have been operational with upgraded
detectors. Coupled with recent improvements in the Tevatron accelerator
performance, the experiments have started producing important physics
results and measurements. 
We report these measurements as well as prospects in the near future.
}

\vspace*{5cm}
\begin{center}

Plenary talk presented at \\ 
\vspace*{5mm}
SUSY 2004 \\

The 12th International Conference   

 on Supersymmetry 
 
and Unification of Fundamental Interactions \\

\vspace*{5mm}

June 17 - 23, 2004 \\

\vspace*{5mm}

Epochal Tsukuba, Tsukuba, Ibaraki, Japan.

\end{center}

\clearpage
This is a blank page.

\clearpage


  \pagestyle{plain}

  \pagenumbering{arabic}

\title{
Electroweak, Top and Bottom Physics at the Tevatron
}

\author{ FUMIHIKO UKEGAWA  (CDF Collaboration)}

\address{ 
Institute of Physics, University of Tsukuba \\
Tennoudai 1-1-1,  Tsukuba-shi,  Ibaraki-ken 305-8571, Japan
\\ {\rm E-mail: ukegawa@hep.px.tsukuba.ac.jp
 \\ {\mbox {}} \\ representing the CDF and D0 collaborations }}

\abstract{
The Tevatron Run-II program has been in progress since 2001,
and the CDF and D0 experiments have been operational with upgraded
detectors. Coupled with recent improvements in the Tevatron accelerator
performance, the experiments have started producing important physics
results and measurements. 
We report these measurements as well as prospects in the near future.
}

\normalsize\baselineskip=15pt

\section{Introduction}

The Tevatron Run-II program officially started in March 2001 
after the previous run (Run I)
ended in 1996. Between these years, the Tevatron accelerator and the CDF and D0
detectors have undergone vast upgrades. The accelerator complex
has added 
the 
Main Injector, replacing the old Main Ring, to inject higher intensity beams 
to the Tevatron and to produce more anti-protons to be used for collisions.
Also the Tevatron beam  energy has been increased 
and it resulted in a center-of-mass energy of 1.96 TeV.
The instantaneous luminosity has improved steadily since the beginning of Run II, 
and at the time of the Conference a record value was $8.3 \times 10^{31}$~cm$^{-2}$~s$^{-1}$.
This is about 5 times higher than the Run-I record value, and almost matches the Run-IIa goal
of $8.6 \times 10^{31}$. 
The integrated luminosity delivered to each experiment 
has exceeded 500~pb$^{-1}$, 
and with about 80\% of them recorded by the detector.

Both the CDF and D0 experiments have broad physics program being carried out
with the data. In the remainder of this manuscript we summarize those in the
areas of electroweak physics, top quark physics and bottom quark physics.
Exotic physics including higgs and susy searches 
is covered by another speaker~\cite{d0-sp}.

\section{Electroweak Physics}
\subsection{Production of single gauge bosons}

\begin{figure}[htb]
\begin{center}
\includegraphics*[width=0.46\textwidth]{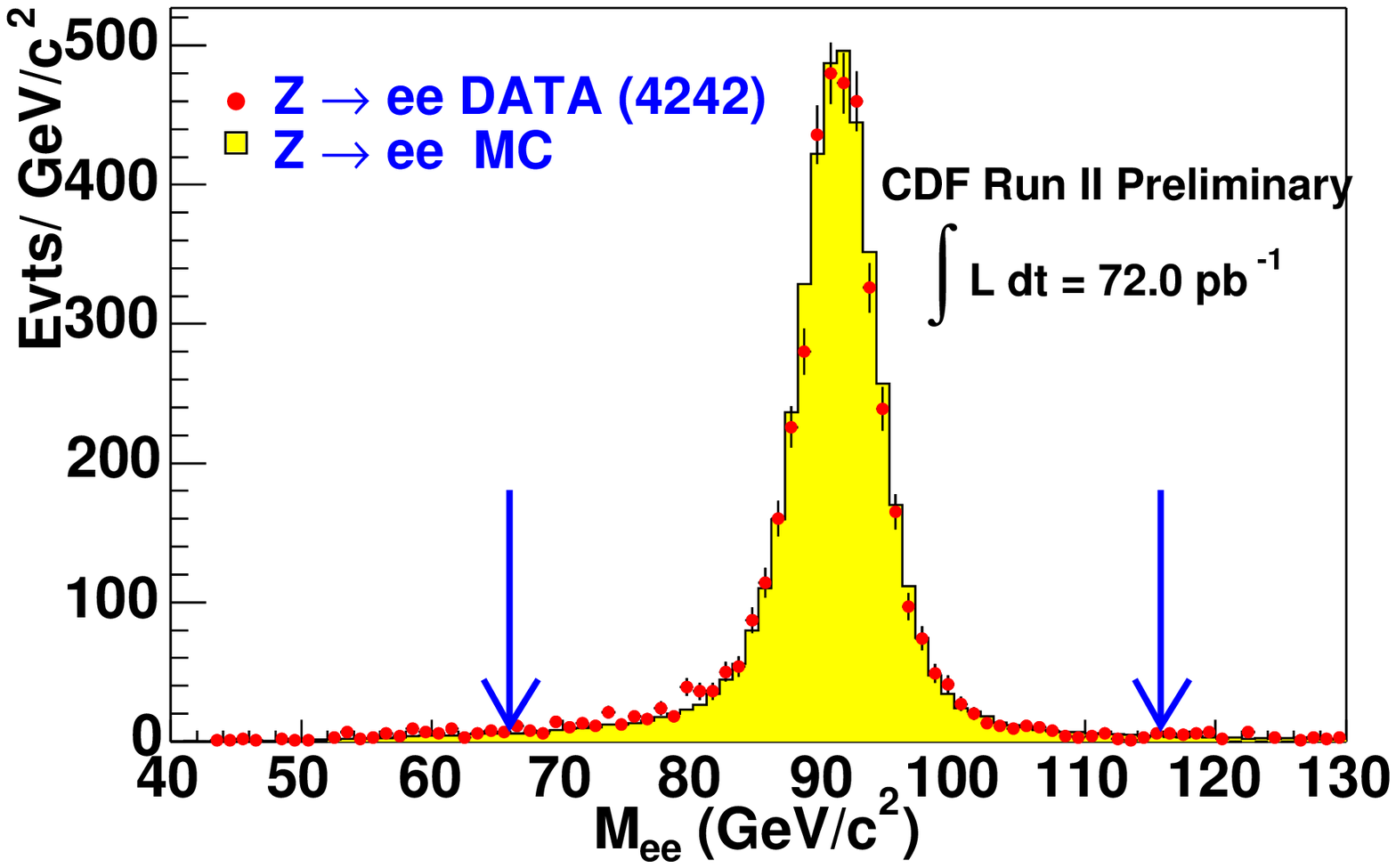}
\includegraphics*[width=0.42\textwidth]{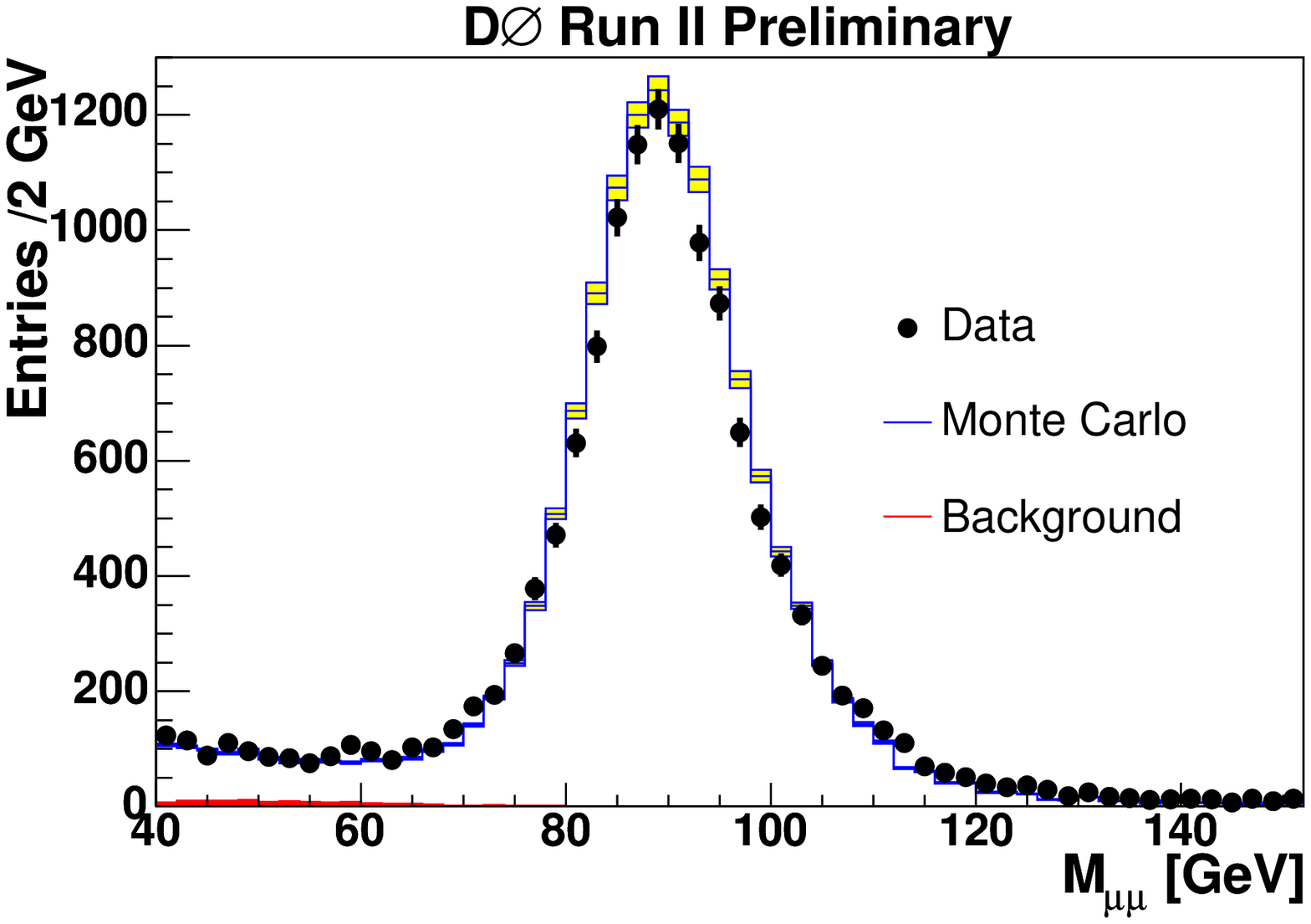}
\caption{%
Invariant mass distributions of lepton pairs
for $Z^0 \rightarrow e^+ e^-$  (left, CDF) and
$Z^0 \rightarrow \mu^+ \mu^-$ (right, D0) candidates.
}
\label{fig:z0}
\end{center}
\end{figure}

\begin{figure}[htb]
\begin{center}
\includegraphics*[width=0.44\textwidth]{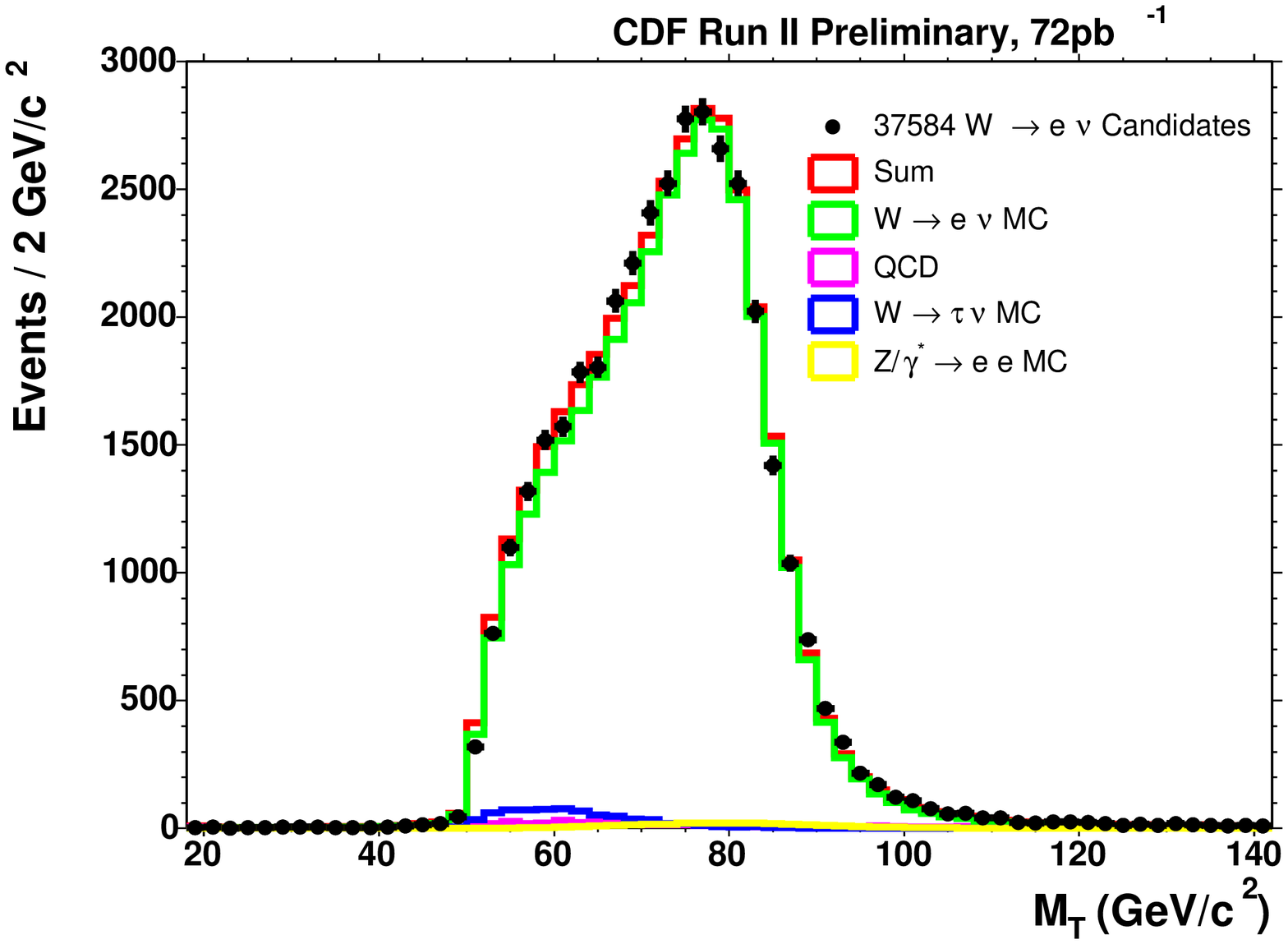} 
\hspace*{3mm}
\includegraphics*[width=0.37\textwidth]{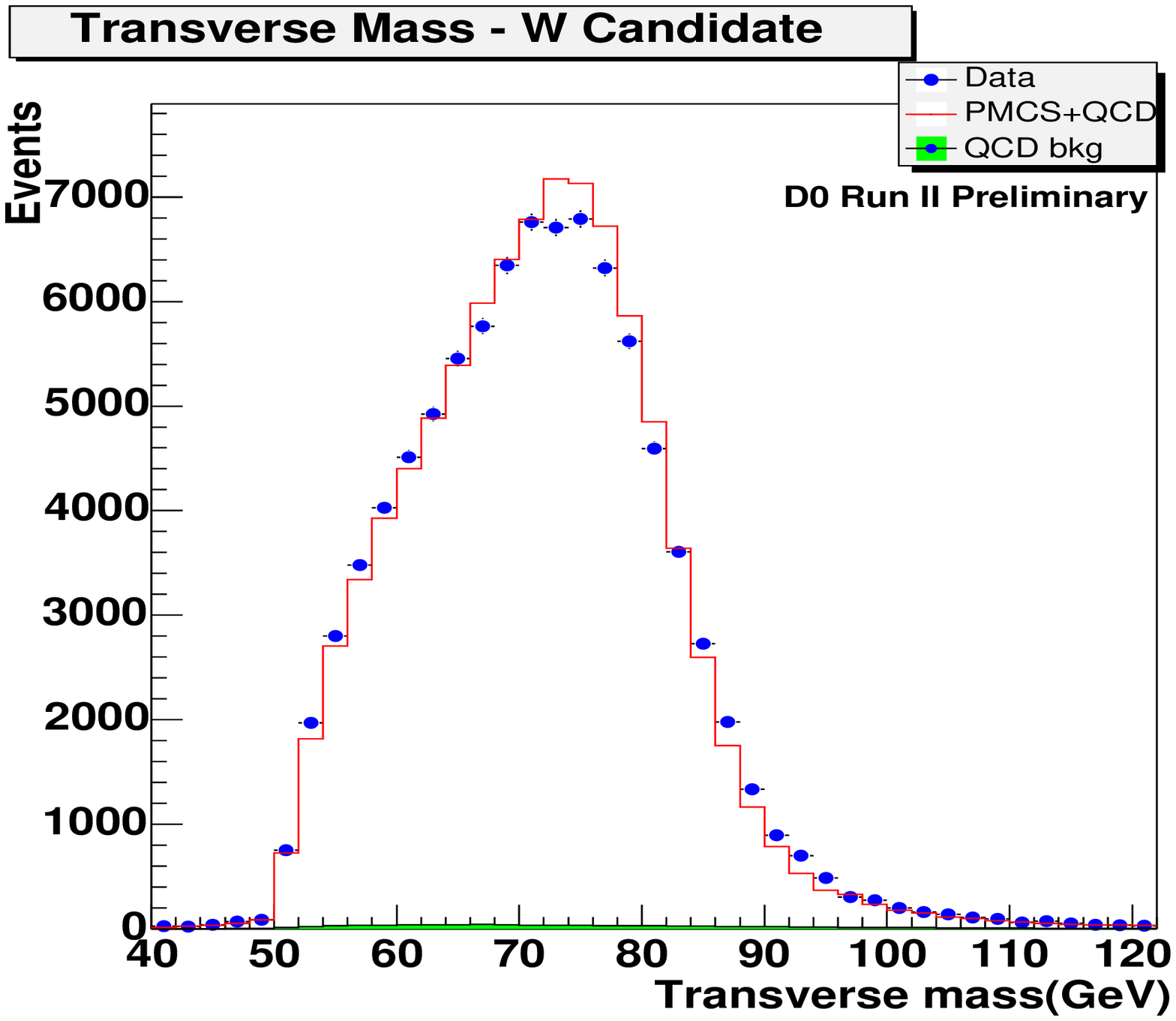}
\caption{%
Transverse mass distributions of lepton and missing $E_T$ system
for $W \rightarrow e \nu$  candidates (left: CDF, right: D0).
}
\label{fig:w-mt}
\end{center}
\end{figure}

\begin{figure}[htb]
\begin{center}
\includegraphics*[width=0.55\textwidth]{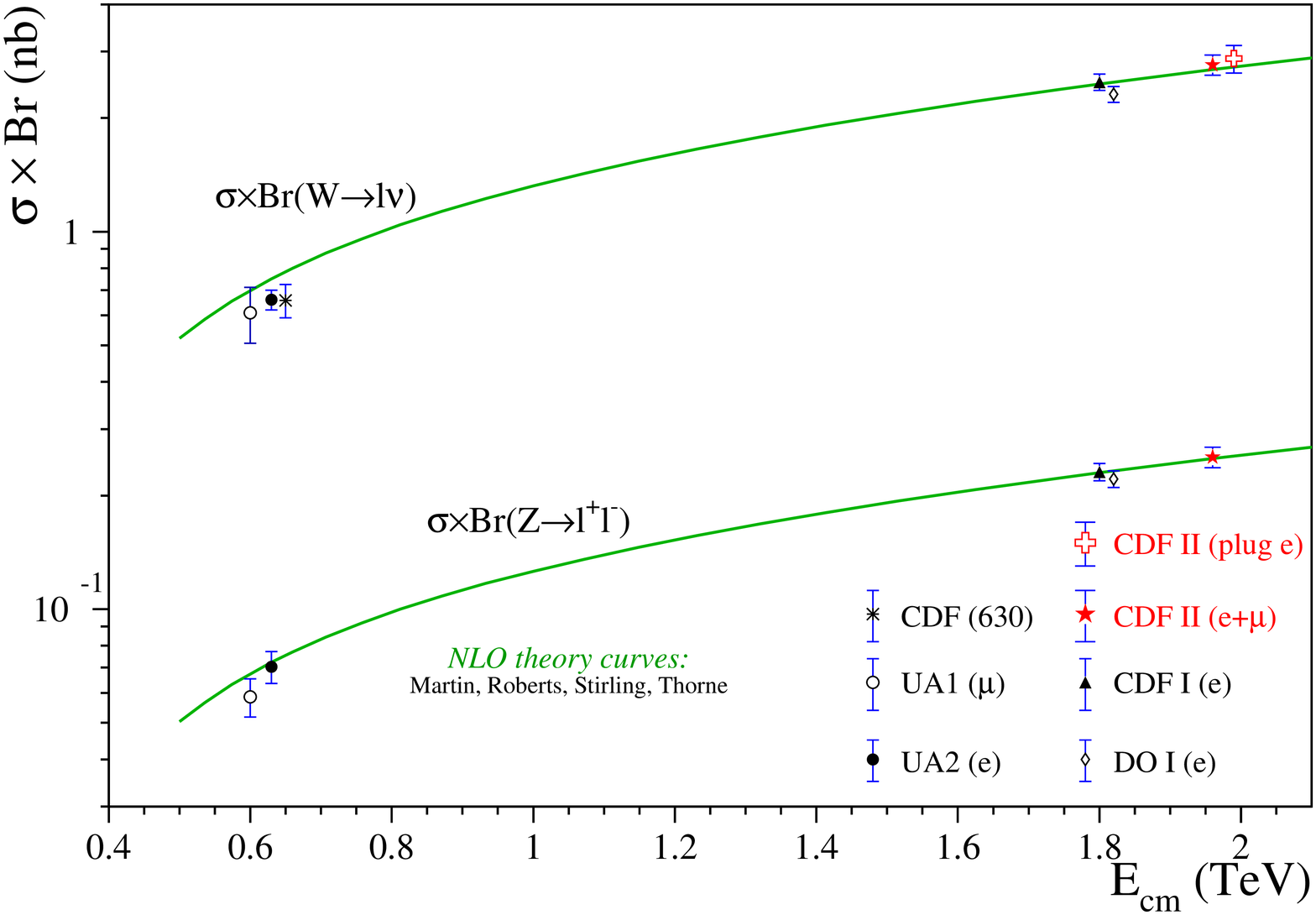} \hspace*{3mm}
\includegraphics*[width=0.29\textwidth]{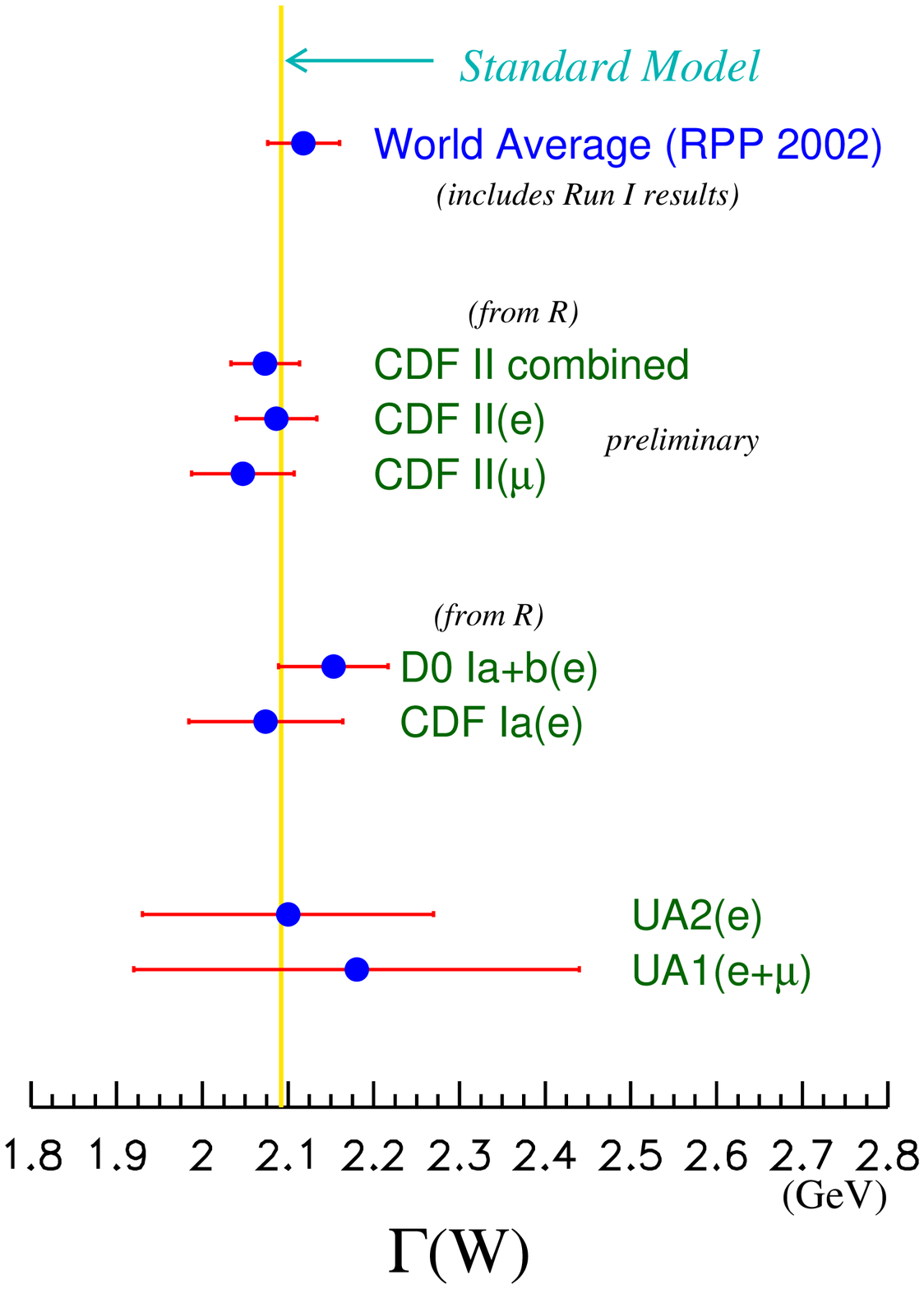}
\caption{%
Left: Production cross section of weak vector bosons
as a function of collision center-of-mass energy.
Right: $W$ boson width measurements.
}
\label{fig:w-z-cross}
\end{center}
\end{figure}

Since the termination of the operation of the LEP2 collider, 
Tevatron is the only accelerator capable of producing $W^\pm$ and $Z^0$ bosons.
CDF and D0 collaborations have performed studies of various aspects
of the weak boson properties. 
They are cleanly identified with their decays to 
leptons (electrons or muons). Figure~\ref{fig:z0} shows invariant mass spectra
of di-lepton candidates from CDF and D0. The production cross sections are 
measured to be~\cite{cdf-z-cross}
\begin{eqnarray*}
\sigma (\bar p p \rightarrow 
Z^0 \rightarrow \ell^+ \ell^- )  & = & 
254.3 \pm 3.3 \pm 4.3 \pm 15.3\ {\rm pb \ (CDF)} \\
& = &
 291.3 \pm 3.0 \pm 6.9 \pm 18.9 \ {\rm pb \ (D0)}, 
\end{eqnarray*}
in good agreement with a theory prediction of 
$250.5 \pm 3.8$~pb (NNLO, MRST)~\cite{w-cross}. 

$W$ boson decays are identified
with an energetic lepton and a large missing transverse energy.
Figure~\ref{fig:w-mt} shows transverse mass distributions. 
The production cross sections are measured to be~\cite{cdf-z-cross,d0-w-cross}
\begin{eqnarray*}
\sigma (\bar p p \rightarrow 
W \rightarrow \ell \nu  )  & = & 
2777 \pm 10 \pm \ \, 52 \pm 167 \ {\rm pb \ \ (CDF)} \\
& = & 
   2865.2 \pm 8.3  \pm 62.8  \pm 40.4  \pm 186.2  \ {\rm pb \ \ (D0 \ }e)\\
& = & 
3226 \pm 128 \pm 100 \pm 322 \ {\rm pb \ \ (D0 \ \  \mu)}, 
\end{eqnarray*}
again in good agreement with theory, 
$2687 \pm 40 \ {\rm pb  \ (NNLO, MRST)}$~\cite{w-cross}.

Figure~\ref{fig:w-z-cross}~(left) shows a summary of those measurements,
along with earlier measurements at the CERN collider, as a function of the 
collision center-of-mass energy.

The ratio $R$ of the $W$ and $Z$ boson production rates, 
defined by
\[ R \equiv 
\frac { \sigma (\bar p p \rightarrow W^\pm ) 
        \cdot 
        {\cal B} (W^\pm \rightarrow \ell \nu ) }
      { \sigma (\bar p p \rightarrow Z^0 )
        \cdot 
        {\cal B} (Z^0 \rightarrow \ell^+ \ell^- )} ,
 \]
includes the branching fraction of the $W$ boson. Using a theory prediction
for the ratio of production cross sections and  measurements  of 
${\cal B} (Z \rightarrow \ell^+ \ell^- )$, 
one can extract 
${\cal B} (W^+ \rightarrow \ell^+ \nu )$, or
the total width assuming
the leptonic partial width. 
The extracted numbers are
$\Gamma_W = 2.071 \pm 0.040$~GeV (CDF) and
$ 2.187 \pm 0.128$~GeV (D0).
They are shown in Figure~\ref{fig:w-z-cross}~(right).


\subsection{Pair production  of gauge bosons}

\begin{figure}[t]
\begin{center}
\includegraphics*[width=0.43\textwidth]{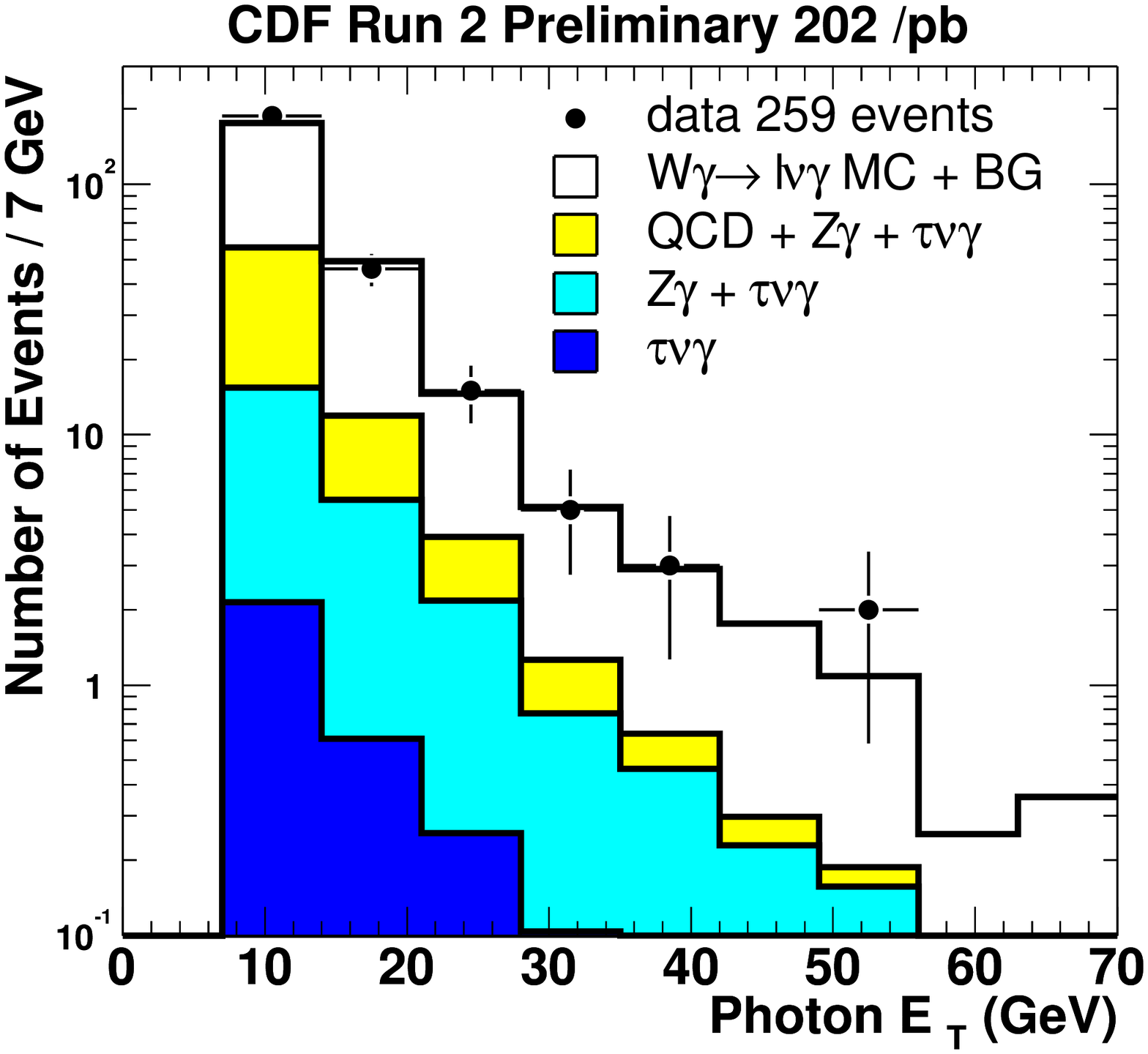} \hspace*{8mm}
\includegraphics*[width=0.4\textwidth]{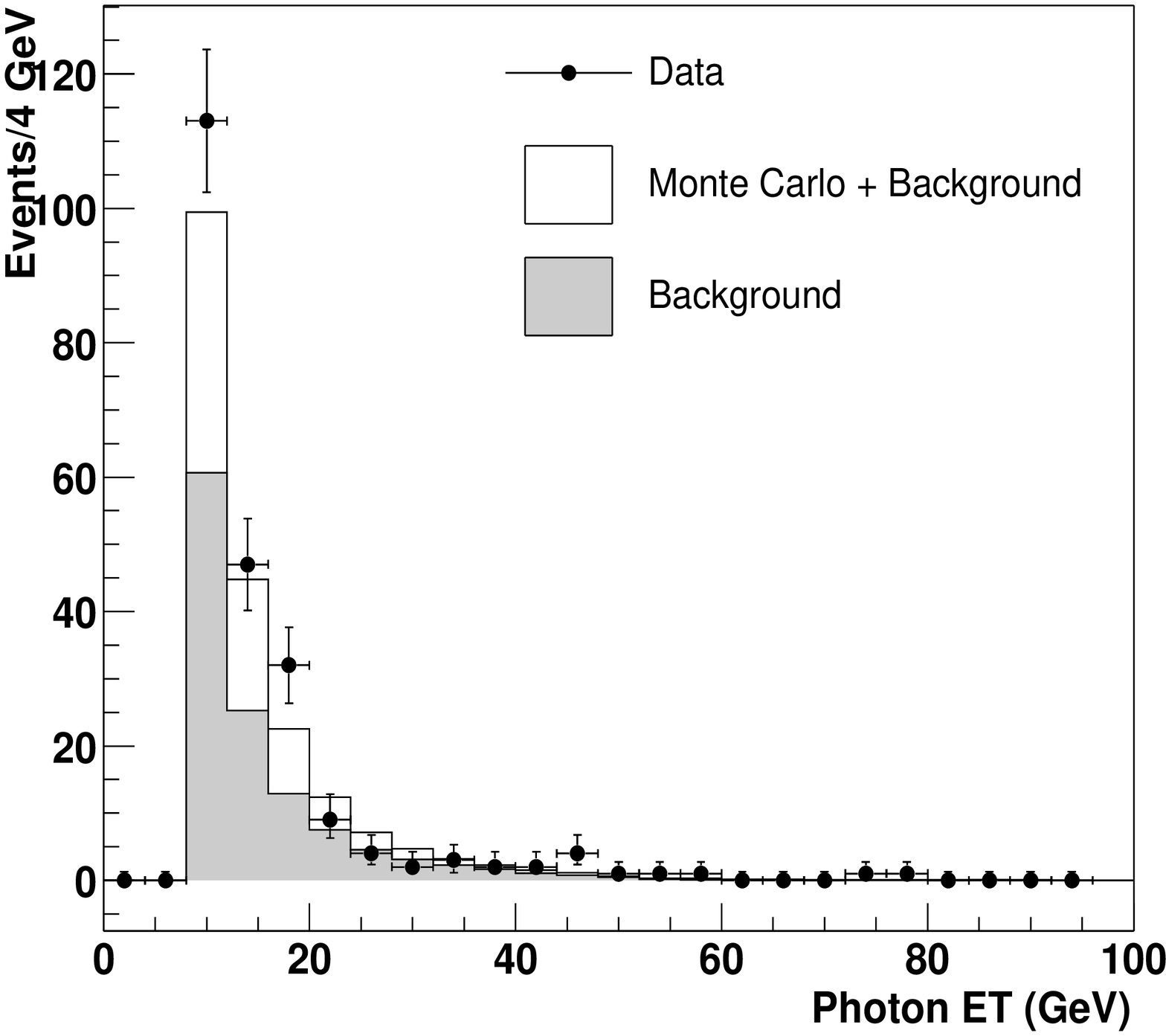}
\caption{%
Transverse momentum distribution of photons produced in association
with the $W$ boson. Left: CDF, right: D0.
}
\label{fig:w-gam}
\end{center}
\end{figure}

The unified electroweak theory has a non-Abelian gauge structure 
and results in self-couplings of gauge bosons. 
CDF has observed the production of $W^+ W^-$ pairs 
for the first time
in hadron colliders. 
The leptonic decay channel of the $W$ bosons is used,
resulting in two energetic leptons and a large missing $E_T$, and no extra jet activities.
The production cross section has been measured to be~\cite{cdf-ww}
\[
\sigma (\bar p p \rightarrow 
W^+ W^- )
= 14.3 \, ^{+  \, 5.6} _{ - \, 4.9}  \pm 1.6 \pm 0.9~{\rm pb} ,
\]
to be compared with a theory prediction of $12.5\pm 0.9$~pb~\cite{theory-ww}.

Associated production of a $W$ or $Z$ boson with a photon is also studied.
Photons are identified in the transverse momentum ranges 
above 7 GeV and 8 GeV in CDF and D0, respectively.
Figure~\ref{fig:w-gam} shows transverse momentum spectra of
photons in $W \gamma$ candidate events. 
Production cross section is measured to be~\cite{cdf-wgam}
\begin{eqnarray*}
& & \sigma ( \bar p p  \rightarrow W \gamma ) 
\cdot {\cal B} ( W \rightarrow \ell \nu  ) \\
& = & 19.7 \pm 1.7 \pm 2.0 \pm 1.1 
         \ {\rm pb \ \ (CDF) 
     \ vs.} \ 19.3 \pm 1.4\ {\rm pb}   \ \ {\rm  (theory)  }  \\
& = & 19.3 \pm 2.7 \pm 6.1 \pm 1.2 
         \ {\rm pb \ \ (D0) 
      \ \ \ \, vs. } \ 16.4 \pm 0.4 \ {\rm pb}  \ \ { \rm  (theory) } ,
\end{eqnarray*}
where the difference reflects the different kinematic requirements.
A more direct test 
of the gauge couplings 
can be performed 
if photon angular distributions of those
events
are studied 
and radiation amplitude zero is looked for directly.

\section {Top Quark Physics}
The top quark was discovered by CDF and D0 in  Tevatron Run-I data of about 100~pb$^{-1}$.
Tens of events were reconstructed then, and therefore all measurements were 
statistically limited.  
The expected 20-fold increase in the amount of data in Run~II should allow
more detailed studies of top quark properties.
They include production cross section, mass, production kinematics, 
$t \bar t$ spin correlations, helicity of $W$ bosons produced in top decays,
branching fractions of various decay modes and searches for rare decays including those
beyond the standard model.

\begin{figure}[t]
\begin{center}
\includegraphics*[width=0.45\textwidth]{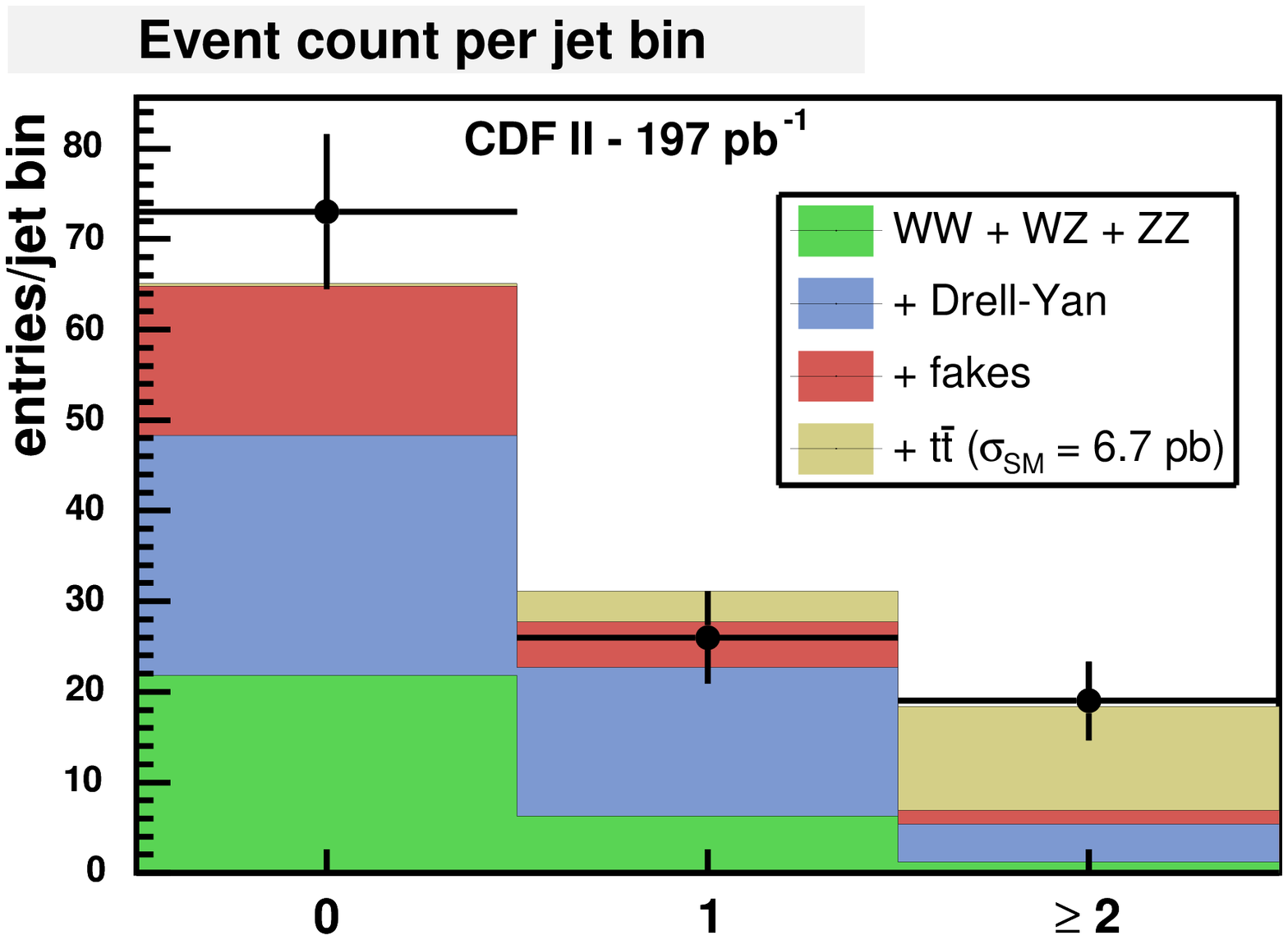} \hspace*{3mm}
\includegraphics*[width=0.38\textwidth]{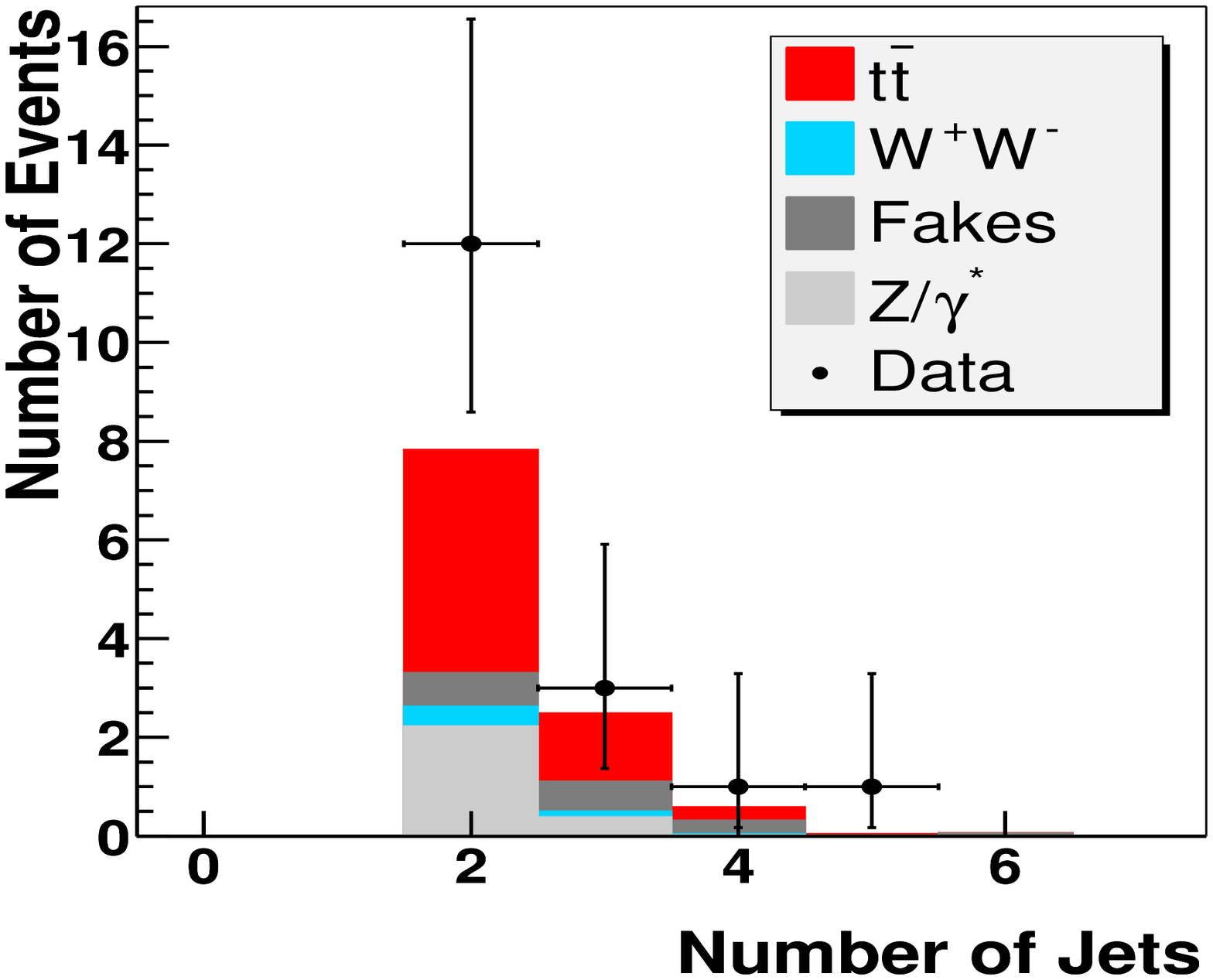}
\caption{%
Jet multiplicity distributions of top candidate events
in the di-lepton channel.
Left: CDF, right: D0.
}
\label{fig:top-jet-mul-dilep}
\end{center}
\end{figure}

\begin{figure}[htb]
\begin{center}
\includegraphics*[width=0.29\textwidth]{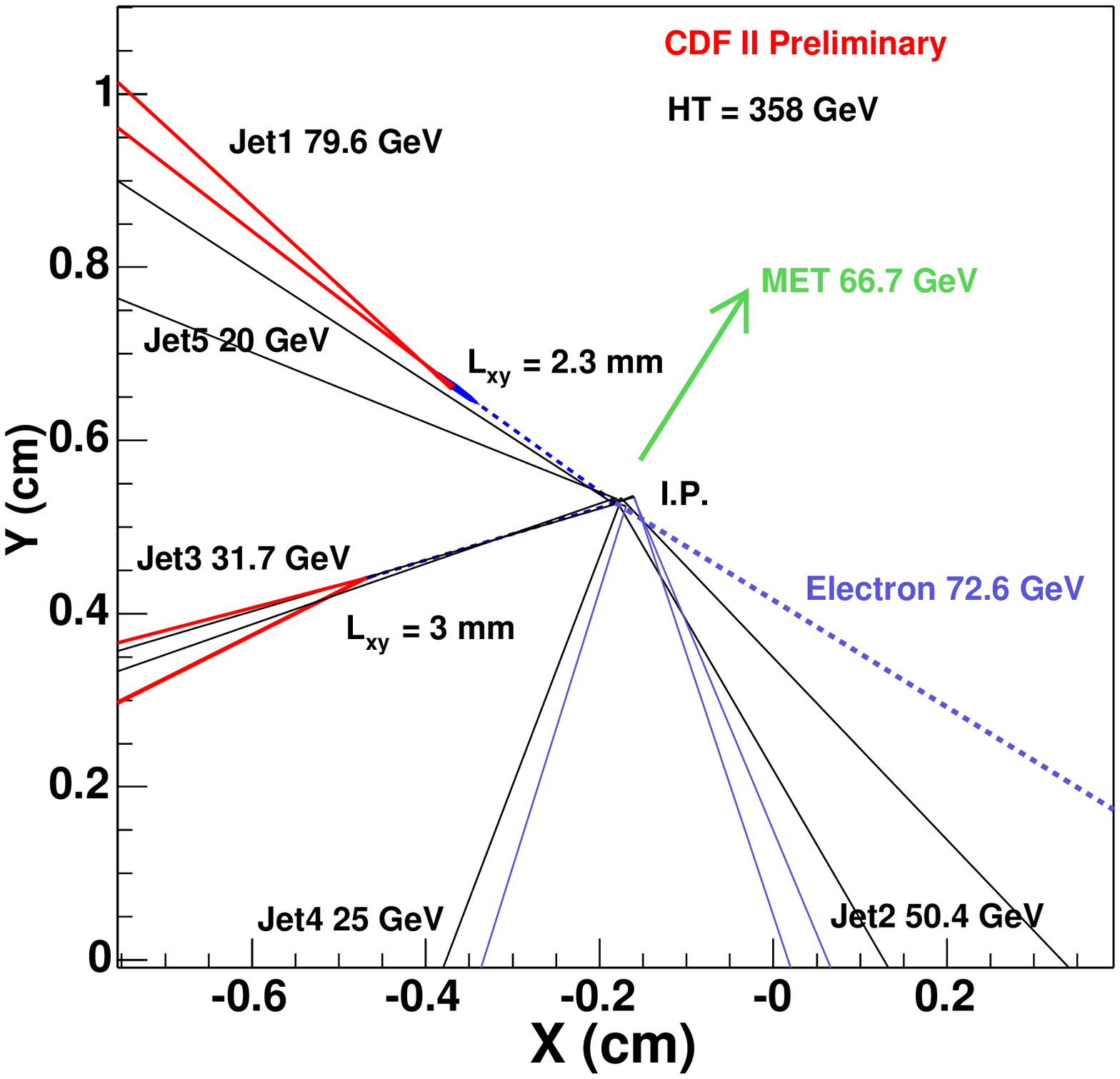} 
\includegraphics*[width=0.35\textwidth]{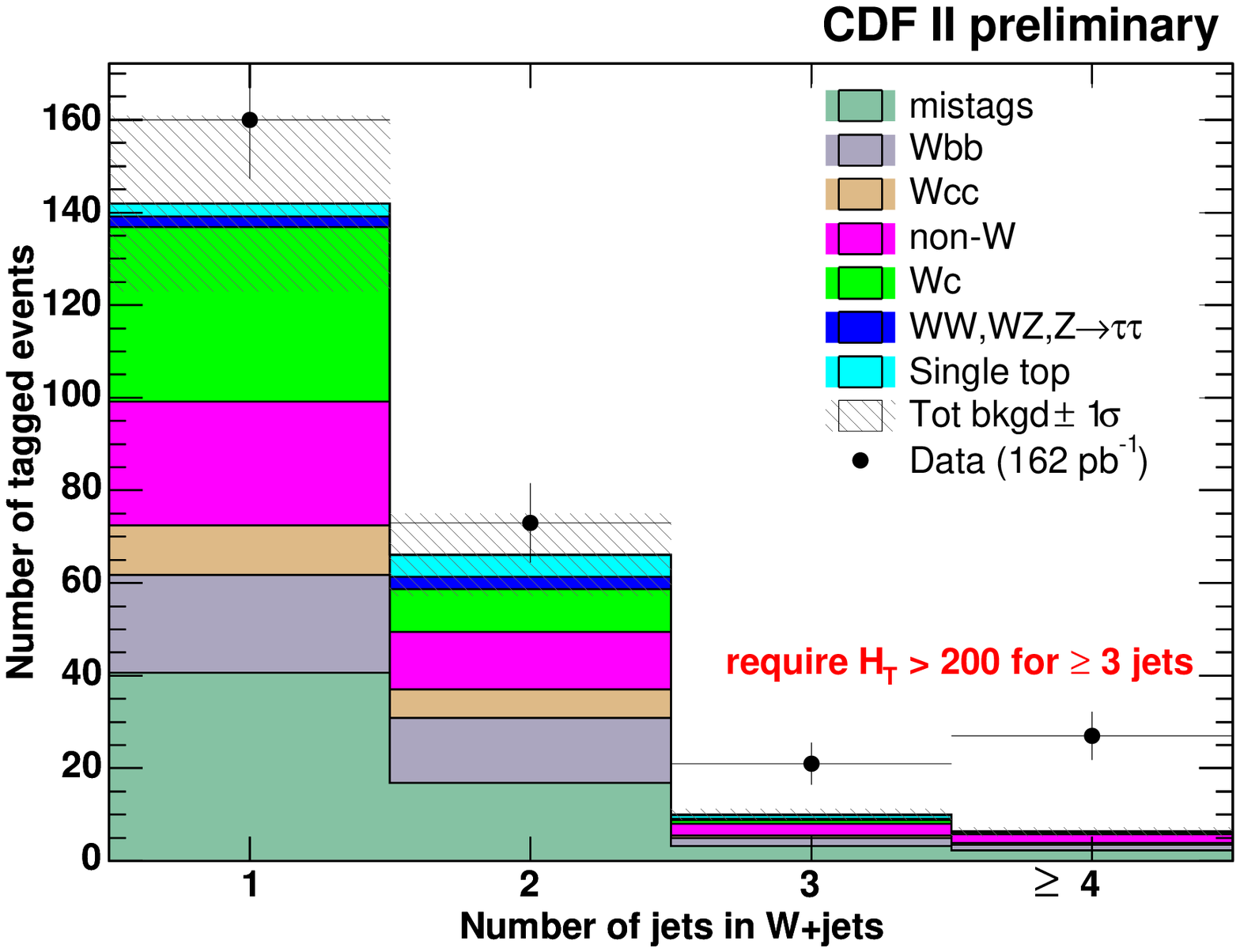}
\includegraphics*[width=0.32\textwidth]{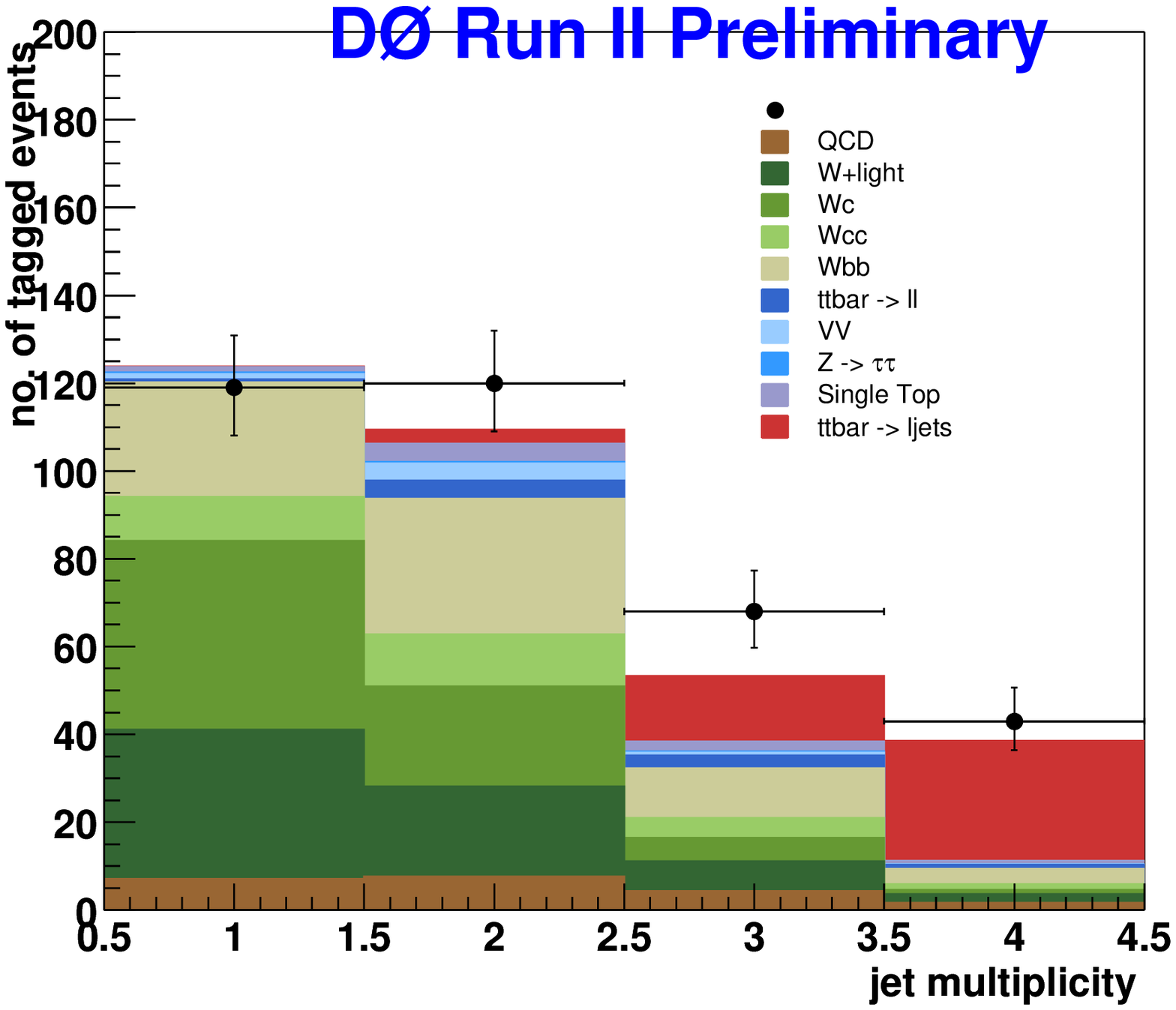}
\caption{%
Left: Example of $b$-tagged top candidate event.
Middle and right: 
Jet multiplicity distributions of top candidate events 
in the lepton plus jet channel from CDF and D0.
The signal region is $N_{\rm jet} \geq 3$.
}
\label{fig:top-example}
\end{center}
\end{figure}

\subsection{Production cross sections}
The signature of top quark pair production and decays is two $W$ bosons and two $b$ quark jets.
Depending on the $W$ decay modes, the final states can be 
two leptons and two jets, one lepton and four jets, 
or six jets. The two $b$ quark jets can be identified
by secondary vertices (reflecting detectable $B$-hadron lifetimes) or ``soft" leptons
from semileptonic decays. 

The di-lepton channel is particularly clean and does not usually
require that the $b$-hjets be identified. 
Figure~\ref{fig:top-jet-mul-dilep} shows the
jet multiplicity distributions of the candidate events in this channel.
The production cross section is measured to be~\cite{cdf-top-cross-dil}
\begin{eqnarray*}
\sigma ( \bar p p \rightarrow  t \bar t X ) 
& = &    \ \, 7.0 \, ^{ +2.7} _{ - \, 2.3} \, ^{ +1.5} _{ - \, 1.3} \pm 0.4 \ 
{\rm pb} \ \ \ {\rm (CDF \ 197  \ pb}^{-1}) \\
& = &  
 14.3 \, ^{ +5.1} _{ - \, 4.3} \, ^{ +2.6} _{ - \, 1.9}   \pm 0.9 \ 
{\rm pb}   \ \ \ {\rm (D0 \ \ \ \ 150  \ pb}^{-1})   .
\end{eqnarray*}

The lepton plus jet final state is less pure, 
and it is required that one or both of $b$-quark
jets be identified. 
An example event with secondary vertex tags is shown in Figure~\ref{fig:top-example}~(left).
The jet multiplicity distributions of these candidate events are
also shown in Figure~\ref{fig:top-example}.
The excess of events in the bins $N_{\rm jet} \geq 3$ bins is nicely described after the
inclusion of top contributions. 
The production cross section is measured to be~\cite{cdf-top-cross}
\begin{eqnarray*}
\sigma ( \bar p p \rightarrow t \bar t X ) 
& = &  \ \,  5.6 \, ^{ +1.2} _{ - \, 1.0} \, ^{ +1.0} _{ - \, 0.7} 
\  \ \ \ \ \ \ \ \,  {\rm pb \ \  \ (CDF \ 162  \ pb}^{-1})                                           \\
 & = & 
   8.2  \pm 1.3    \, ^{ +\, 1.9} _{ - \, 1.6} \pm 0.5
\ {\rm pb \ \ \ 
  (D0  \ \ \  160  \ pb}^{-1})  .
\end{eqnarray*}
There exit many other measurements of the quantity 
using various different techniques~\cite{top-other}.

\subsection{Top quark mass}

\begin{figure}[htb]
\begin{center}
\includegraphics*[width=0.40\textwidth]{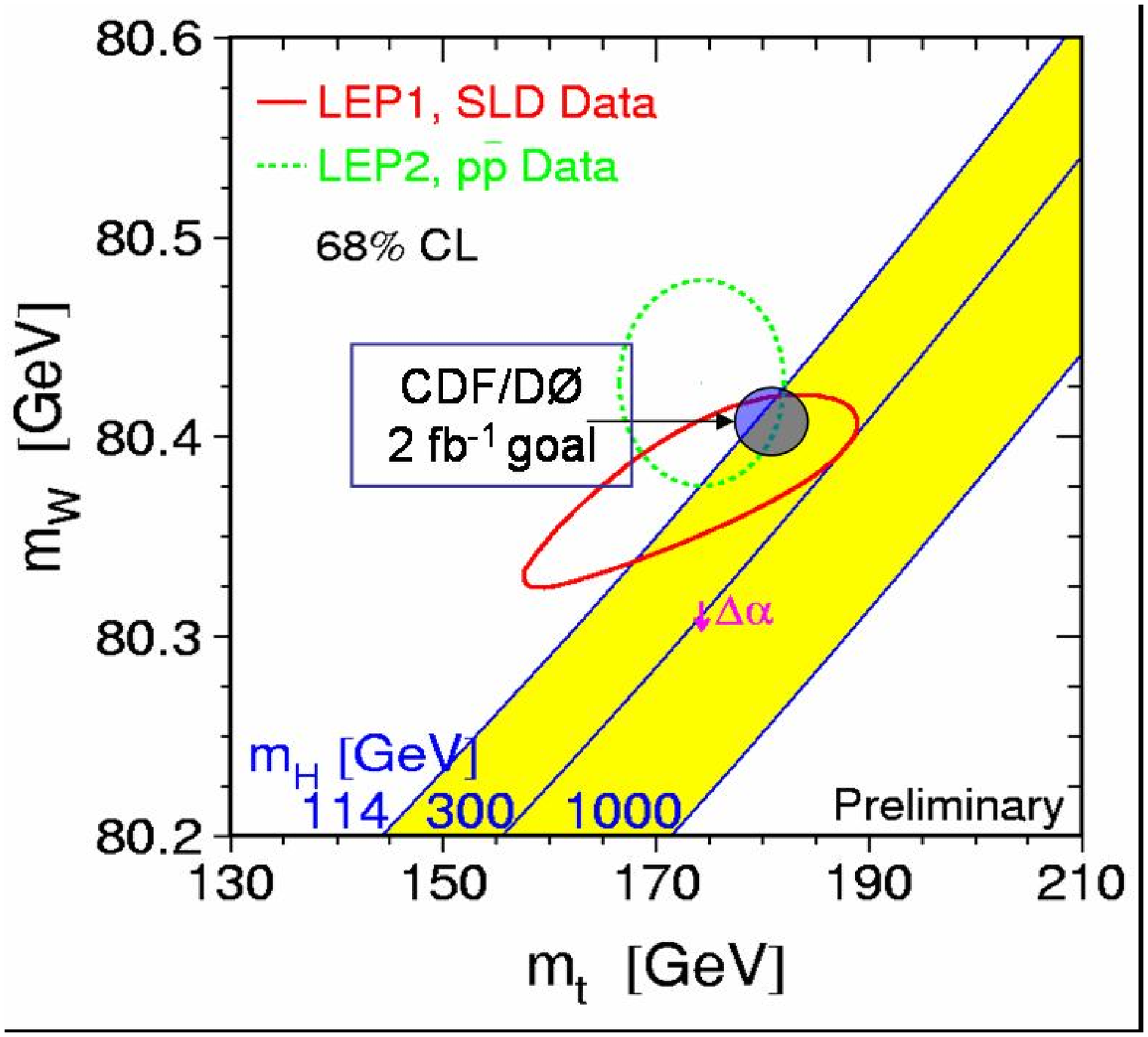} \hspace*{3mm}
\includegraphics*[width=0.45\textwidth]{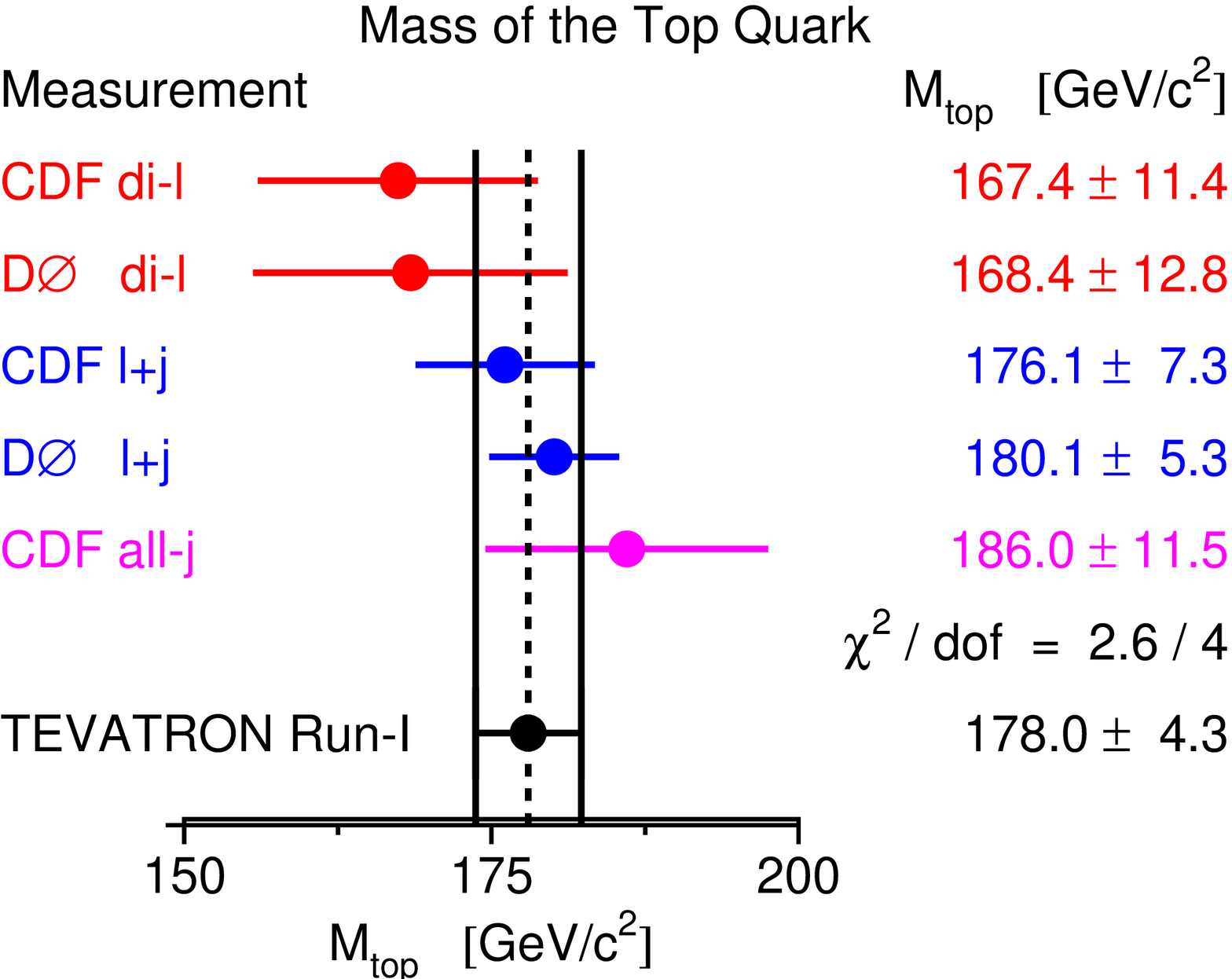}
\caption{%
Left: Top quark and $W$ boson mass measurements
and their relation to the Higgs boson mass.
Right: Summary of top quark mass measurements from Run I.
}
\label{fig:mw-mtop}
\end{center}
\end{figure}

\begin{figure}[htb]
\begin{center}
\includegraphics*[width=0.42\textwidth]{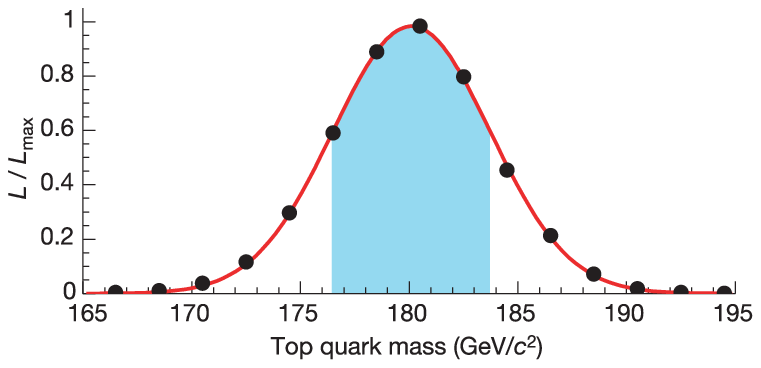}
\hspace*{3mm}
\includegraphics*[width=0.40\textwidth]{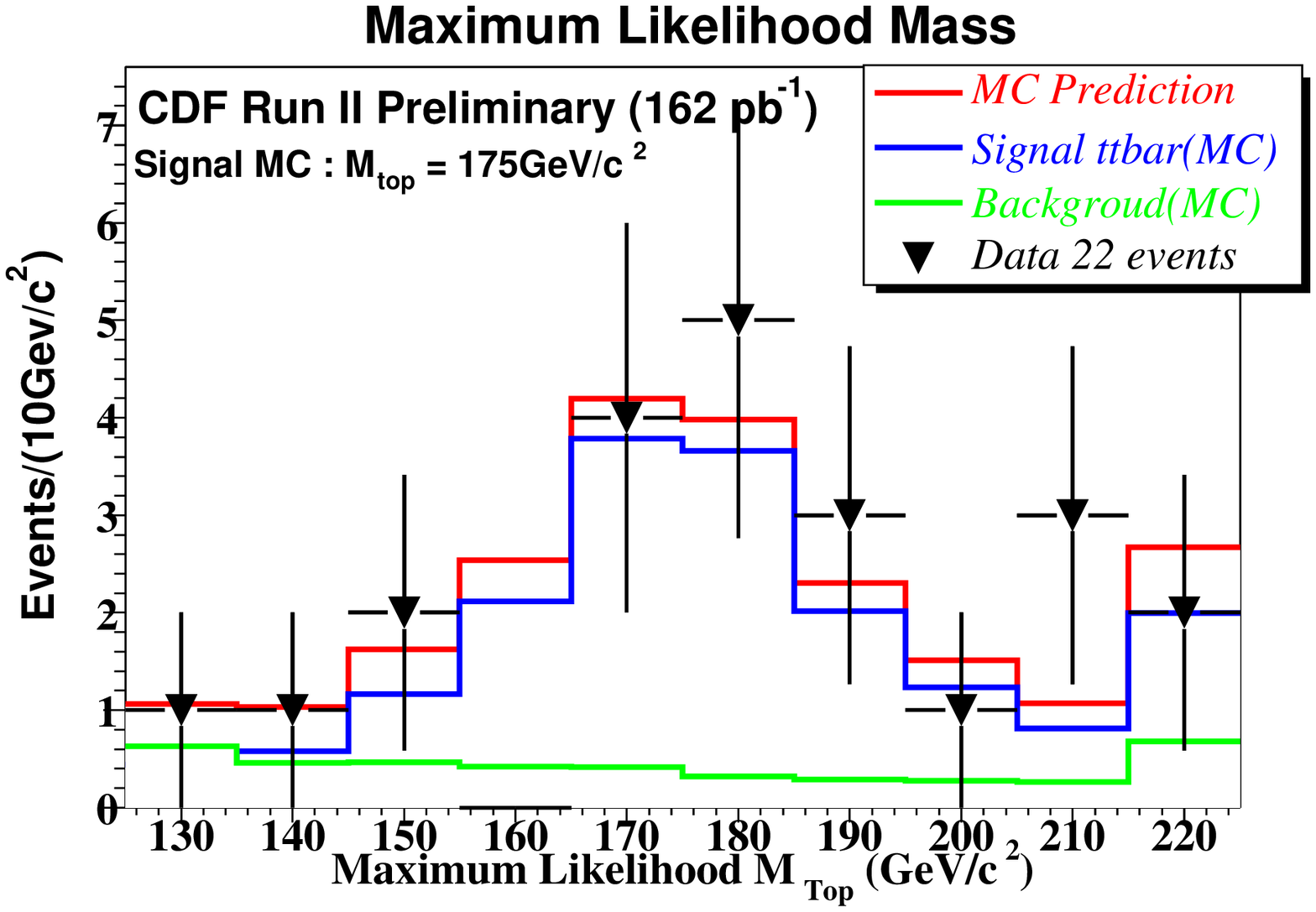} 
\caption{%
Left: Top quark mass likelihood distribution 
from D0 in re-analysis of Run-I data.
Right: Distribution of top quark mass reconstructed with DLM at CDF.
}
\label{fig:top-mass}
\end{center}
\end{figure}

The mass of the top quark is an important quantity to 
measure, not just for its own right, 
but also from other physics perspectives. 
When combined with the $W$ boson mass, 
it can provide indirect information on the Higgs boson mass. 
Figure~\ref{fig:mw-mtop} shows this relation, 
together with an expected precision in the measurements 
of the $W$ boson and top quark masses in Tevatron Run II~\cite{run1-top}. 
The figure also summarizes Run-I measurements of the top quark mass.

The D0 collaboration has re-analyzed Run-I data using a new technique of
determining the top quark mass, 
utilizing maximal information from the events, including
matrix elements for $t\bar t$ production and decay and parton
distributions~\cite{d0-nature}. 
The extracted top quark mass is
\[
m_t = 180.1 \pm 3.6  \pm 4.0 \ {\rm  GeV} / c^2 .
\]
The CDF Collaboration has applied a method called Dynamical Likelihood Method,
originally developed in the '80s~\cite{DLM}, to measure the top quark mass.
Figure~\ref{fig:top-mass} shows the reconstructed top mass
distribution. The extracted mass value is~\cite{mass-dlm}
\[
m_t = 174.9 \, ^{+ \, 4.5} _{- \, 5.0} \pm 6.2 \ {\rm GeV}/ c^2.
\]

There exit many other measurements using various techniques, 
including those made public
since the time of this Conference.
However, we will not describe them in this report. 
We refer the reader to Ref.~\cite{top-other}.

\subsection{$W$ helicity in top decays}

\begin{figure}[t]
\begin{center}
\includegraphics*[width=0.35\textwidth]{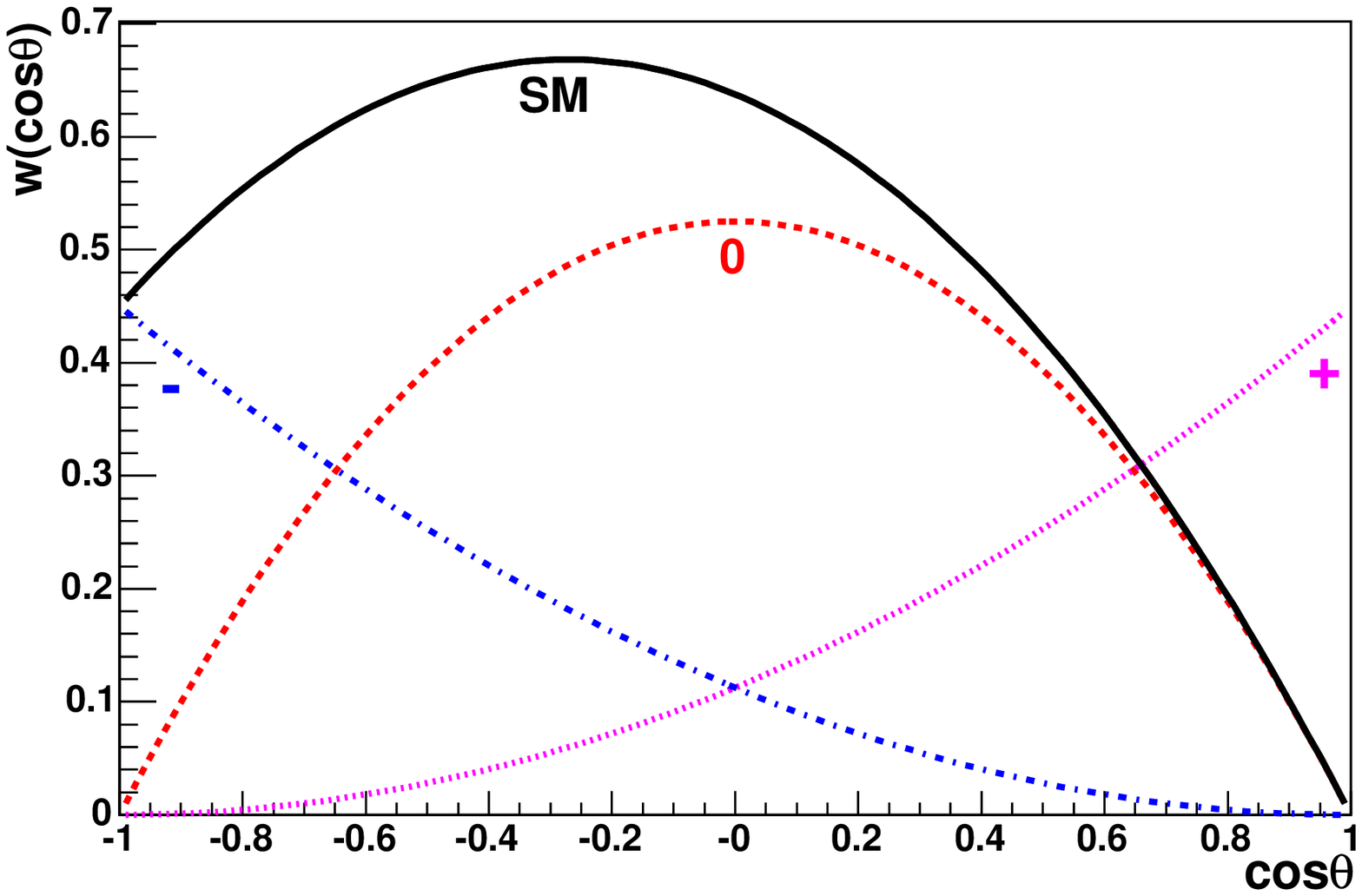} 
\includegraphics*[width=0.29\textwidth]{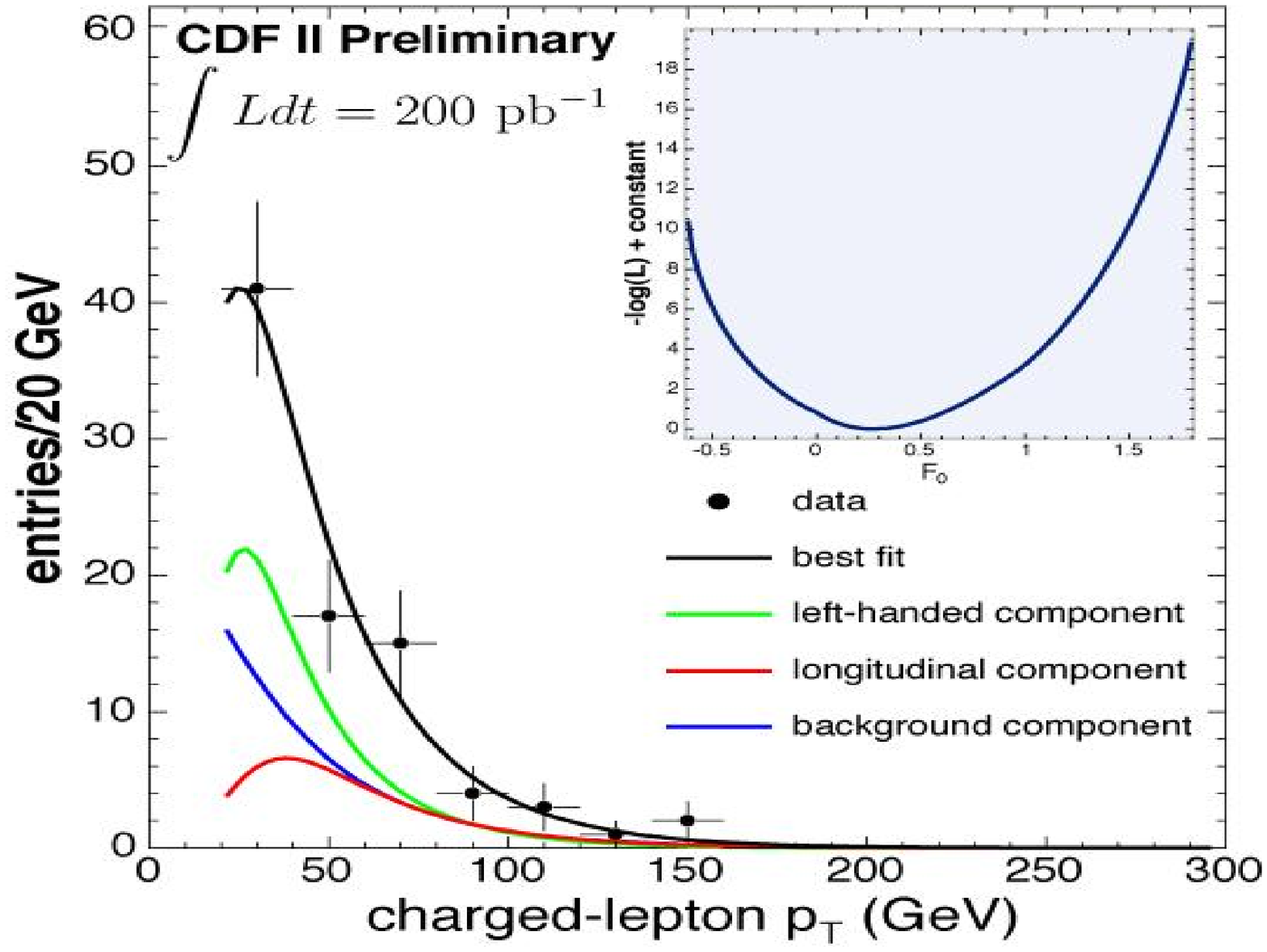}
\includegraphics*[width=0.33\textwidth]{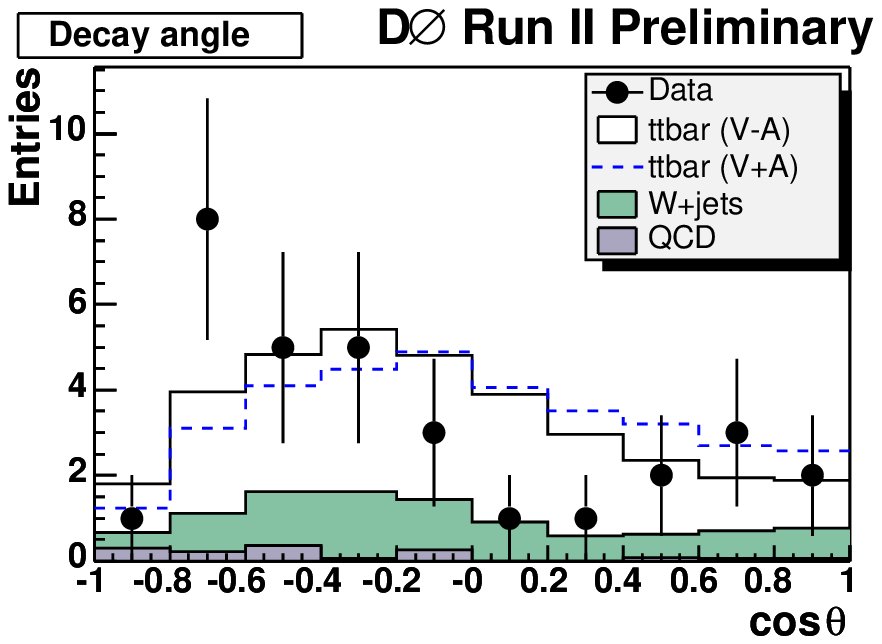}
\caption{%
Left: Lepton angular distributions in the rest frame of the $W$ boson
with respect to the top quark direction
in the cases of  longitudinal $(0)$ and left-handed $(-)$ $W$ bosons.
Middle: Transverse momentum distribution of leptons in top candidate events (CDF).
Right: Lepton angular distribution in lepton$+$jet top candidate events (D0).
}
\label{fig:w-angle}
\end{center}
\end{figure}

The $W^-$ boson produced in the top quark decay can be either left-handed or longitudinally 
polarized. Their mixture is predicted reliably by the standard model and is
\begin{eqnarray*}
f_0 & \equiv &  
\frac { \Gamma (  t \rightarrow b \, W_0 )   }
      {  \Gamma ( t \rightarrow b \, W_0 ) 
       + \Gamma ( t \rightarrow b \, W_L )  } 
= \frac { m_t^2 } { m_t^2 + 2 \, m_W^2} \\
& = & 
0.70 \ \ \ {\rm for } \ \ m_t = 175 \ \ {\rm GeV/}c^2 ,
\end{eqnarray*}
where $f_0$ is the fraction of the longitudinal component. 
Two different methods have been used to extract the fraction.
The first examines the lepton momentum spectrum.
The different $W$ polarization states 
give different lepton angular distributions in the top quark decay.
This is shown in Figure~\ref{fig:w-angle}. 
The angle $\theta^*$ is the angle of the lepton momentum direction
in the rest frame of the $W$ boson 
with respect to the top quark momentum direction.
The left-handed $W$ boson produces leptons 
peaked toward the backward direction, 
while for the longitudinal 
$W$ boson the distribution is symmetric. 
The right-handed $W$ boson would give forward-peaked leptons, 
which is absent in the standard model.

The lepton momentum in the laboratory frame reflects
these angular distributions, and is harder for the leptons from
longitudinal $W$ and softer for those from
the left-handed $W$. The lepton spectrum from CDF is shown
in Figure~\ref{fig:w-angle}~(middle). 
The extracted longitudinal fraction $f_0$,
under the assumption that the right-handed component is absent,
is~\cite{d0-helicity}
\[
f_0 = 0.27 \, ^{ + \, 0.35} _{ - \, 0.24} ,
\]
not inconsistent with the standard model prediction.
D0 reconstructs the angle $\theta^*$ and examines its 
distribution. It is shown in Figure~\ref{fig:w-angle}~(right).
The fraction of the (non-SM) right-handed polarization ($f_+$) 
is fitted for, with the longitudinal fraction $f_0$ fixed to the standard model value, 
and is determined to be~\cite{d0-helicity}
\[
    f_+ = -0.13 \pm 0.23,   \ \ {\rm or} \ \ 
 f_+ < 0.244     \ \ (90\% \ {\rm CL} ) .
\]


\section{Bottom Quark Physics}

Since the confirmation of large CP violation in some of $B$ hadron decay modes a few years ago,
the thrust of $B$ physics is now in  testing the Kobayashi-Maskawa picture of
CP violation and in particular the consistency of the unitarity triangle,
and in searches for possible effects of new physics such as supersymmetry.
A lot of excitement has emerged since the summer of 2003, 
when the Belle Collaboration 
announced a possible hint of new physics 
in a measurement of CP asymmetry in the 
$B^0 \rightarrow \phi K^0_S$ decay. 
Within the standard model, the CP asymmetry measured in this decay mode should be
identical to that measured in the (well-established) 
$B^0 \rightarrow J/\psi K^0_S$ decay mode.
In both cases 
CP asymmetry arises, in the standard model, 
from the complex phase of $B^0 \bar B^0$ mixing 
and is $\sin 2\beta$.
However, if a new phase exists in their decays, 
the asymmetries in the two modes can be different. 
The $B^0 \rightarrow \phi K^0_S$ decay 
proceeds via a quark-level transition $b  \rightarrow s \bar s s$,
which is a loop process in the standard model 
and is suppressed
relative to tree-level processes.
The asymmetry in the $B^0 \rightarrow \phi K^0_S$ mode 
as of summer 2003 was 
$-0.96 \pm 0.51$~\cite{belle}\footnote{
As of ICHEP 2004, the new value of CP asymmetry in 
the $\phi K^0$ mode from Belle 
is  $ +0.06 \pm 0.33 \pm 0.09$ (hep-ex/0409049).
}, 
which is about 3.5~$\sigma$ away
from $+ 0.731 \pm 0.056$ measured with $b \rightarrow c \bar c s$ modes.
Therefore, if new physics exists, it is in the $b \rightarrow s$
transitions.

The CDF and D0 experiments can provide unique tests of some of the  $b \rightarrow s$ 
processes, taking advantage of  decays of the $B^0_s$ meson, 
which cannot be produced at the $\Upsilon(4S)$
resonance.
They are :
\begin{itemize}
\item $B^0_s \bar B^0_s$ oscillations. 

\item Search for CP violation in $B^0_s \rightarrow J/\psi \, \phi$.

\item Measurement of CP asymmetries in $B^0_d \rightarrow \pi^+ \pi^-$
and $B^0_s \rightarrow K^+ K^-$ modes.

\item Search for rare decays $B^0_{s,d} \rightarrow \mu^+ \mu^-$.

\end{itemize}
We discuss each of them in some detail below. 
Before doing that, however, we describe
a  benchmark $B$ physics measurement 
from D0, which concerns the ratio of the charged and neutral $B$ mesons,
$B^-$ and $\bar B^0$.

\begin{figure}[htb]
\begin{center}
\includegraphics*[width=0.45\textwidth]{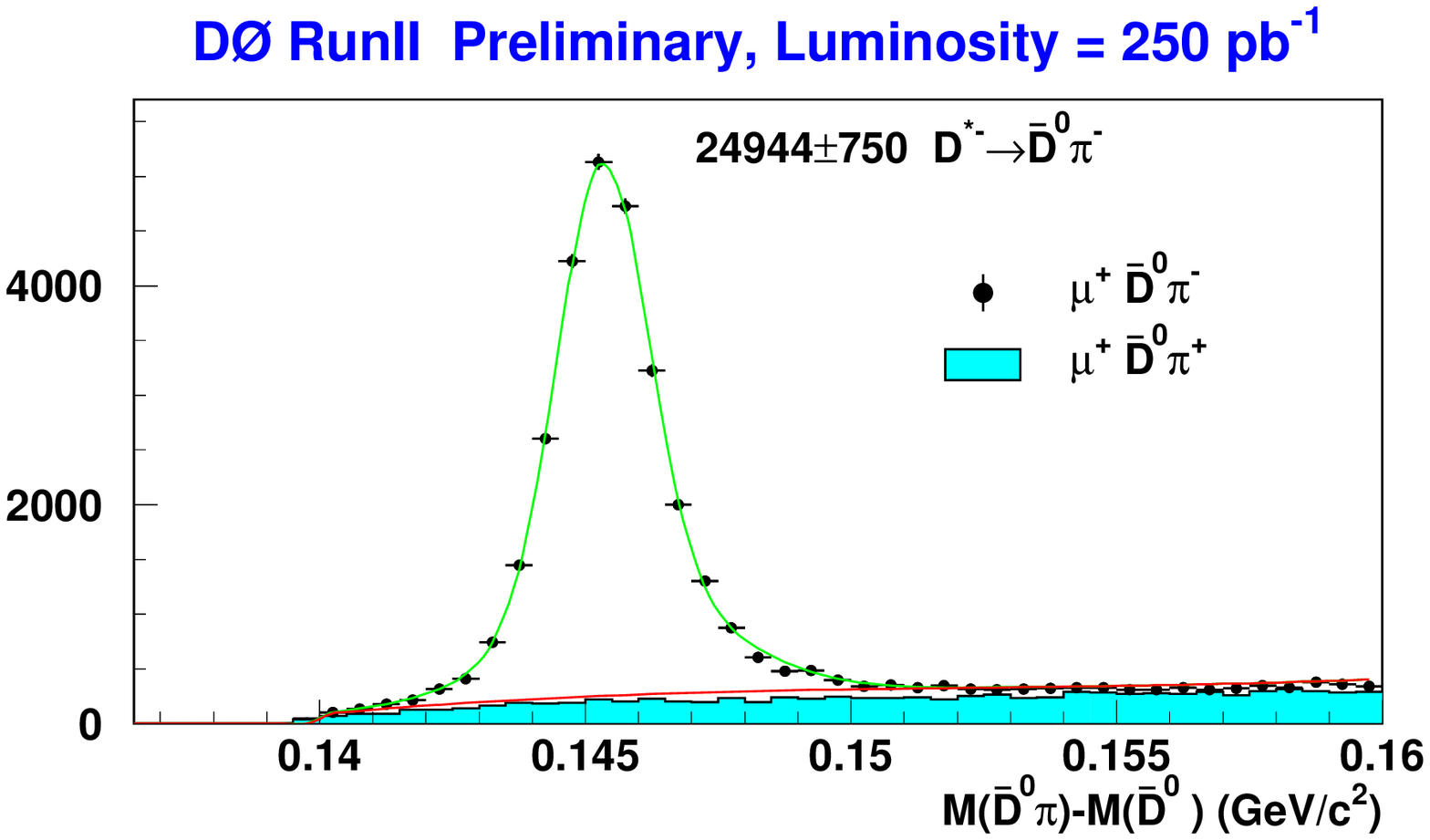} 
\hspace*{3mm}
\includegraphics*[width=0.45\textwidth]{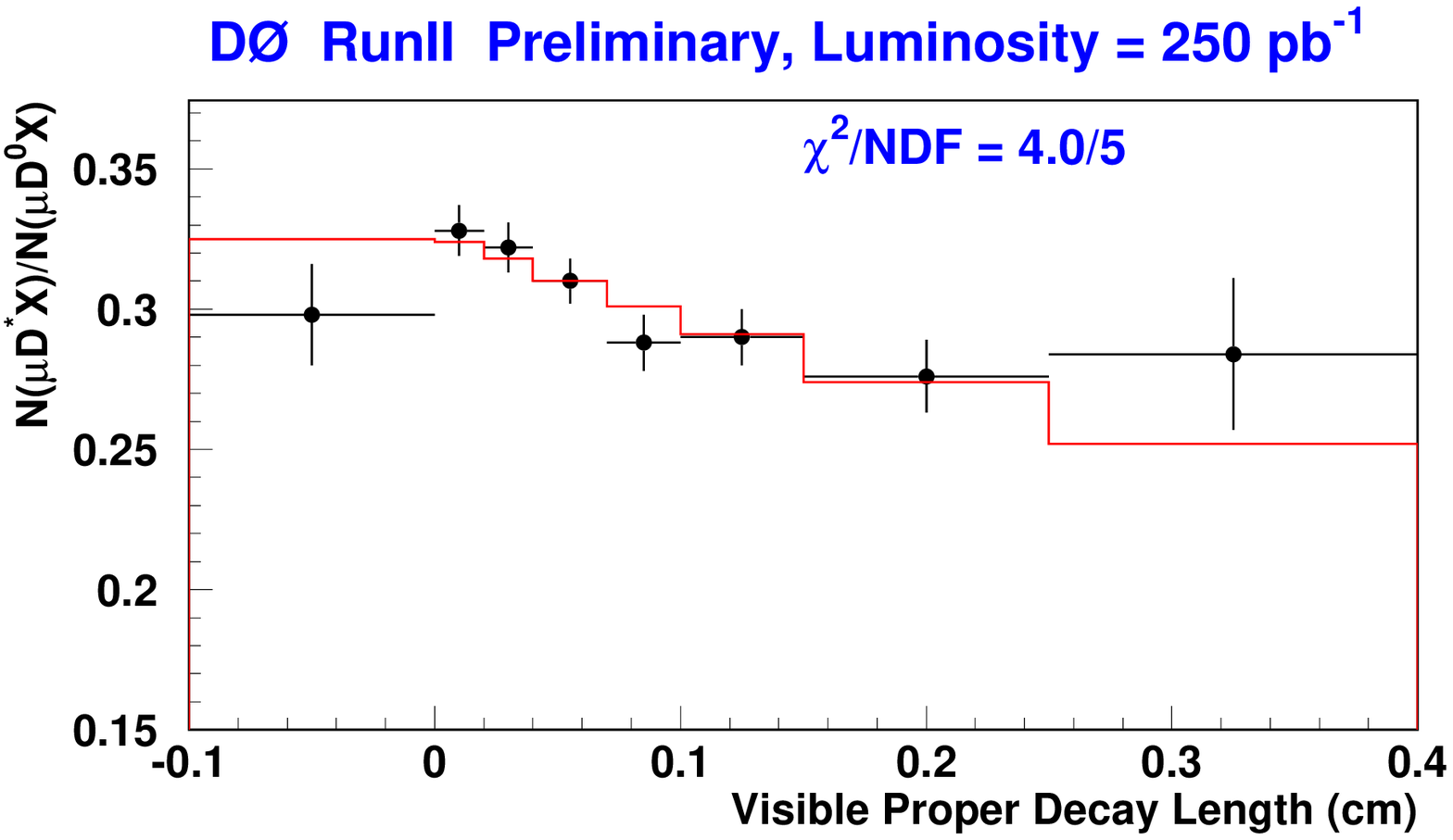}
\caption{%
Left: Signal of $\bar B \rightarrow \mu^- \bar \nu D^{*+} X$ decays
reconstructed in D0 data.
Right: Ratio of $\mu^- D^{*+} $ and $\mu^- D^0$ yields
as a function of their estimated proper decay time.
}
\label{fig:d0-dstar}
\end{center}
\end{figure}

The lifetimes of different $B$-hadron species 
are of interest,
because they offer probes into $B$-hadron decay mechanisms
beyond the simple spectator model picture.
One way to measure their lifetimes separately is
to use signals of fully reconstructed decays.
Another way is to use semileptonic decays, which can be written as
$\bar B \rightarrow \ell^- \bar \nu {\bf D}$, where
${\bf D}$ is a charm hadron system whose charge is
correlated with the parent $B$ hadron charge. 
To be more specific, 
the $\ell^- D^{*+}$ final state is dominated by the $\bar B^0$ meson decays, 
and 
the $\ell^- D^0$ final state is dominated by the $B^-$ meson decays
(provided that those coming from $D^{*+}$ decays are excluded),
allowing us to extract the two lifetimes.
The D0 signal of the $\bar B \rightarrow \mu^- \bar \nu D^{*+} X$ decay
is shown in Figure~\ref{fig:d0-dstar}~(left). 
D0 examines the lifetime dependence of 
the ratio  of the rates of the two final states,
$\mu^- D^{*+}$ and $\mu^- D^0$. 
If the lifetimes of the two parent $B$ meson
states are different, the ratio should change as a function
of the decay time. 
Figure~\ref{fig:d0-dstar}~(right) shows this dependence,
and the ratio clearly deceases with an increasing decay time,
meaning that the $\bar B^0$ meson has 
a shorter lifetime than the $B^-$ meson.
The extracted number is~\cite{d0-life}
\[
  \tau(B^-)  / \tau( \bar B^0 )  = 1.093 \pm 0.021 \pm 0.022,
\]
consistent with recent measurements at the $B$ factory and other experiments.

\begin{figure}[t]
\begin{center}
\includegraphics*[width=0.43\textwidth]{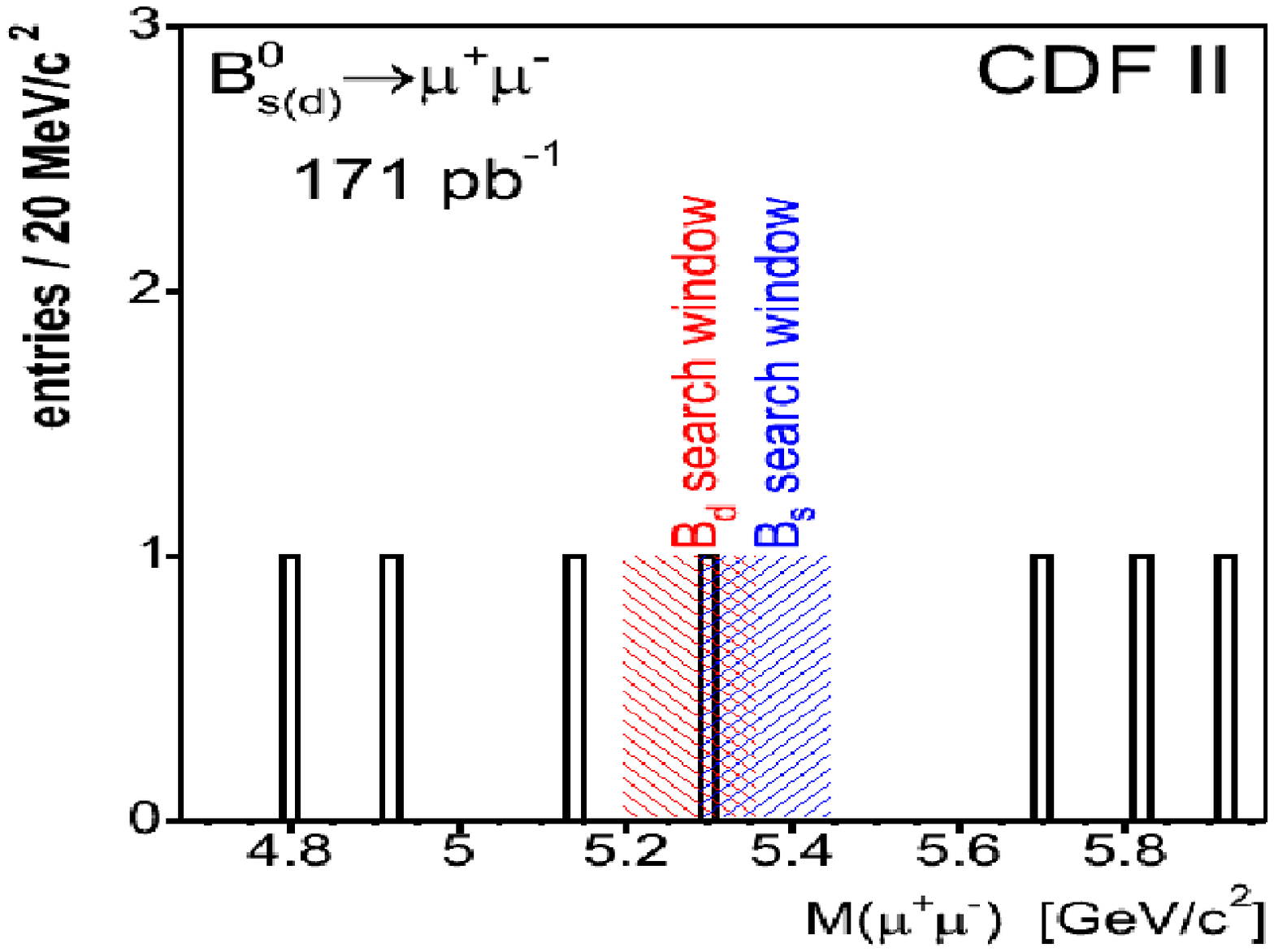}
\hspace*{3mm}
\includegraphics*[width=0.48\textwidth]{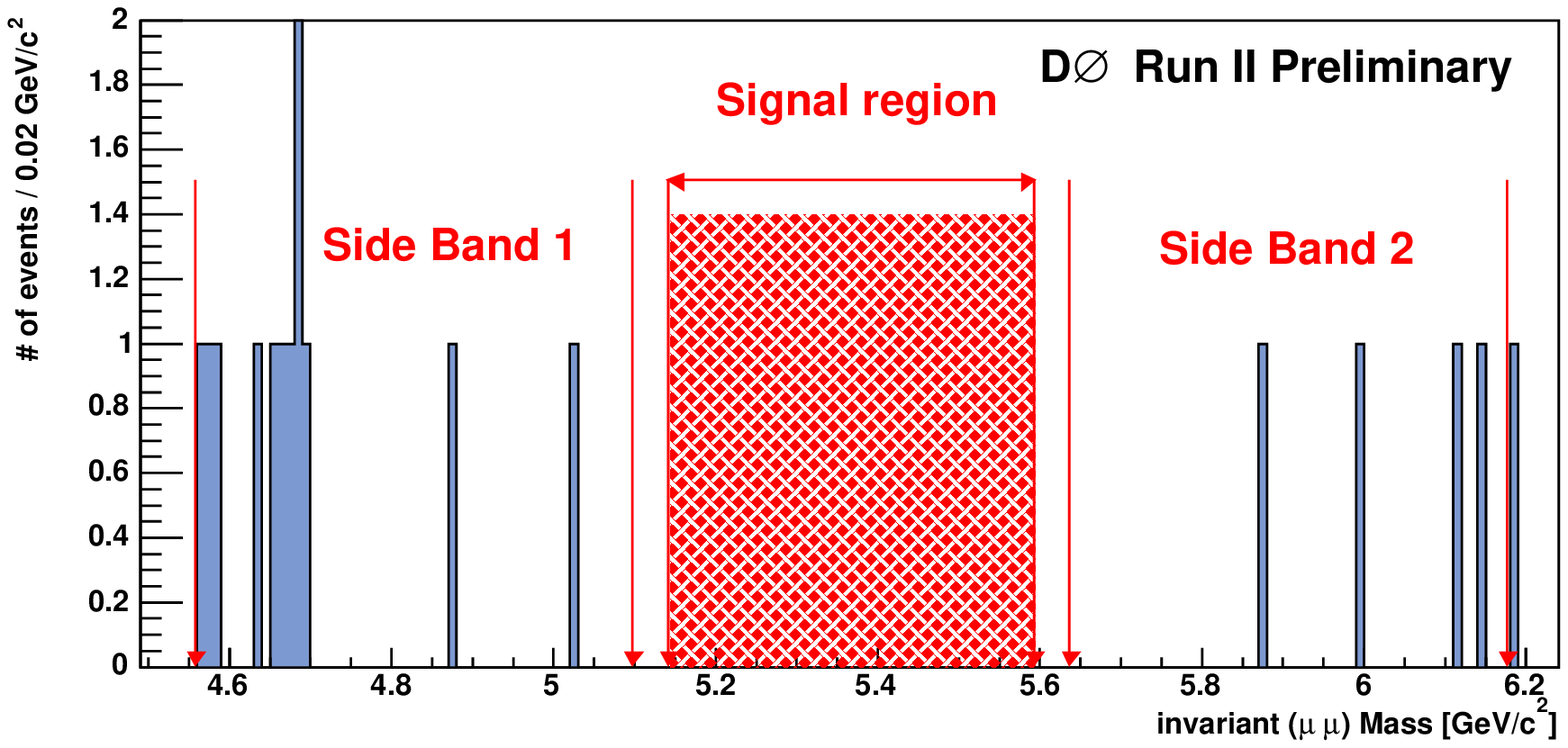}
\caption{%
Dimuon invariant mass distributions near the $B$ meson mass.
Left from CDF and right from D0.
}
\label{fig:bmumu}
\end{center}
\end{figure}

\vspace*{5mm}
Now we discuss $b \rightarrow s$ flavor-changing processes 
and related topics
that can be studied at the Tevatron.

\subsection{Search for rare decays $B^0_{s,d} \rightarrow \mu^+ \mu^-$}
In the standard model these decays can proceed 
via higher-order, box and loop,
diagrams with weak bosons in the intermediate states. 
They are also suppressed by CKM factors, $| V_{ts} | ^2 $ for the $B^0_s$ meson
and $ | V_{td} | ^2  $ for the $B^0_d$ meson. 
Furthermore, the initial state is spin zero,
so they are suppressed by helicity conservation.
The standard model predictions for the branching fractions are~\cite{buras}
\begin{eqnarray*} 
{\cal B } (  B^0_s  \rightarrow \mu^+ \mu^-   ) & = & 
 (3.4 \ \pm  \ 0.5  \ ) \times 10^{- \, 9}    \\
{\cal B } (  B^0_d  \rightarrow \mu^+ \mu^-   )   & = & 
 (1.00 \pm 0.14 ) \times 10^{-10} .
\end{eqnarray*}
The values for the corresponding electron modes are five orders of 
magnitude smaller. 
They are extremely small values, 
and so the decay is a good place to look for 
effects of new physics.

Both the CDF and D0 experiments have performed the search. 
Figure~\ref{fig:bmumu}~(left) shows an invariant mass spectrum
of CDF dimuon candidate events near the $B$ meson mass.
The shaded regions show the search windows. 
One candidate event 
is found in the overlap region of $B_d^0$ and $B_s^0$ mesons
in a data sample of 171~pb$^{-1}$, 
while the expected number of background events is $1.1 \pm 0.3$.
The following upper limits (95\% CL) 
have been placed~\cite{bmumu}
\begin{eqnarray*}
{\cal B} (  B^0_s  \rightarrow \mu^+ \mu^-  )   & <  & 
                7.5 \times 10^{-7}    \\
{ \cal B} (  B^0_d  \rightarrow \mu^+ \mu^-  )  & < & 
  1.9 \times 10^{-7} ,
\end{eqnarray*}
which improves the previous CDF limits by a factor of three.
The mass distribution from D0 is also shown in the figure. At the time of
the Conference they had completed a sensitivity study but 
had not opened the signal region data box yet. 
The estimated sensitivity
with 180~pb$^{-1}$ of data 
was 
 $  {\cal B}(B^0_s \rightarrow \mu^+ \mu^-) 
 \sim  10.1 \times 10^{-7} 
 $
at the 95\% CL. 
After the Conference D0 has improved the sensitivity further and 
obtained a 95\% CL upper limit of~\cite{d0-bsmumu-pub}
\[
{\cal B}(B^0_s \rightarrow \mu^+ \mu^-) 
 <   5.0 
\times 10^{-7}  .
 \]

\begin{figure}[t]
\begin{center}
\includegraphics*[width=0.45\textwidth]{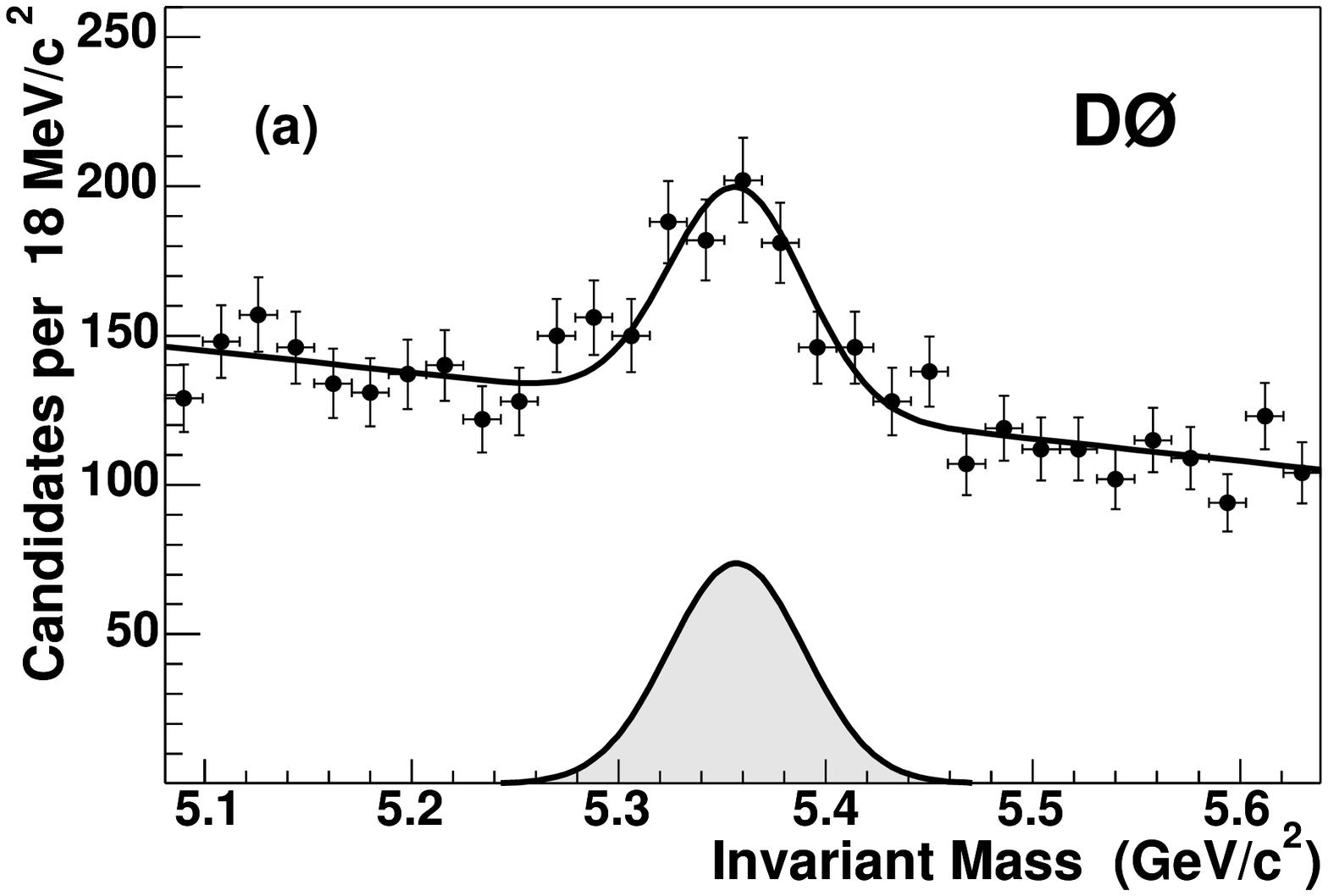} 
\hspace*{3mm}
\includegraphics*[width=0.40\textwidth]{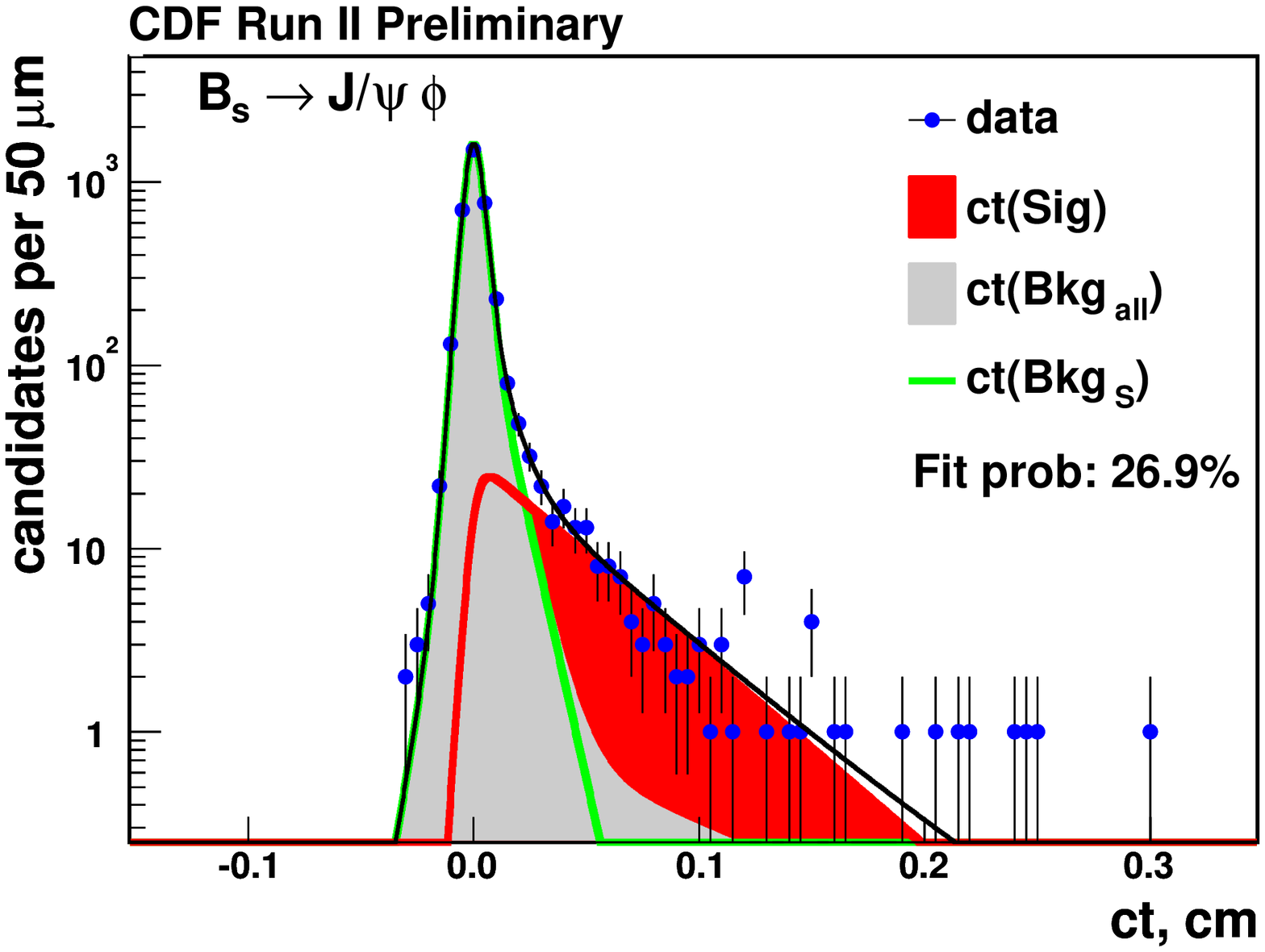}
\caption{%
$B^0_s \rightarrow J/\psi \phi$ decays reconstructed 
(left, D0) and the decay time distribution (right, CDF).
}
\label{fig:bs-psi-phi}
\end{center}
\end{figure}

\subsection{Studies of $B^0_s \rightarrow J/\psi \, \phi$ decays}

The decay mode $B^0_s \rightarrow J/\psi \, \phi$ is 
attractive experimentally because
it provides distinctive signatures. 
The reconstruction can be done with relative ease.
Figure~\ref{fig:bs-psi-phi}~(left) shows the signal from the
D0 experiment in 220~pb$^{-1}$ of data.
The lifetime of the $B^0_s$ meson is measured to be~\cite{d0-bslife}
\[
    \tau(B^0_s) = 
  1.473 \, ^{ + \, 0.052 } _ {-\, 0.050 }  \pm 0.023 \ {\rm ps} . 
\]
CDF reconstructs a similar signal (not shown), whose decay time
distribution is shown in Figure\ref{fig:bs-psi-phi}~(right). CDF extracts~\cite{cdf-bs-mass-life}
\begin{eqnarray*}
  m(B^0_s) & = & 5366.01 \pm 0.73 \pm 0.33 \ {\rm  MeV}/ c^2  \\ 
  \tau(B^0_s) & = & 1.369 \pm 0.100 \pm 0.010  \ {\rm ps} 
\end{eqnarray*}
using a data sample of 240~pb$^{-1}$.

Theory predicts that the lifetime of the $B^0_s$ meson
should be the same as that of the $B^0_d$ meson within 
1\%.
However, a sizable width difference, 
$\Delta \Gamma_s / \Gamma_s$ of order 10\%~\cite{lenz}, 
can exist 
between the two mass eigenstates of the $B^0_s \bar B^0_s$ system.
The decay mode currently in discussion has been measured to be
dominated by a CP-even state, 
which should correspond roughly to 
the shorter lifetime mass eigenstate
of the two~\cite{buchalla}. 
Therefore, the lifetime measured with this mode can be shorter if 
it is compared to other measurements of the $B^0_s$ meson lifetime, 
which come mostly from flavor-specific final states, 
or if it is compared to the $B^0_d$ meson lifetime.

As for the CP content of the decay mode, 
CDF has measured the polarizations in the decay as well as in the
$B^0 \rightarrow J/\psi K^{*0}$ mode.
The fraction of the transverse helicity state (CP odd) 
is measured to be~\cite{dgog}
\begin{eqnarray*}
\Gamma_\perp / \Gamma &  = &  0.183 \pm 0.051 \pm 0.054 
\ \ \ ( B^0_d \rightarrow J/\psi \,  K^{*0} )   \\
 & = & 0.232 \pm 0.100 \pm 0.013 \ \ \ ( B^0_s \rightarrow J/\psi \, \phi ) .
\end{eqnarray*}
After the Conference, CDF released~\cite{dgog} a result of 
an attempt to measure $\Delta \Gamma_s$, 
which gives a value of
\[
    \Delta \Gamma_s  / \Gamma_s = 0.65 \, ^{+0.25} _ {-0.33} \pm 0.01 , 
\]
though it is not significant yet.
A sizable $\Delta \Gamma$ is important also because it may
allow CP studies of $B^0_s$ meson decays without the necessity
of flavor tagging.

In the future, this decay mode will be used to search for
mixing-induced CP violation.  
The mode is a $B_s^0$ equivalent
of the $B^0_d \rightarrow J/\psi K^0_S$ mode, except that the
$B^0_s$ mode is not a pure CP eigenstate. 
If the phase of particle-antiparticle oscillations 
is non-zero, it can lead to CP asymmetry in time-dependent decay rates
modulated with the oscillation frequency $\Delta m$.
In the standard model, 
$B^0_s \bar B^0_s$ mixing receives 
very little complex phase because it is
arg($V_{ts}$). Therefore, if CP violation is found to be sizable,
it will be an unambiguous signal of new physics.

\begin{figure}[t]
\begin{center}
\includegraphics*[width=0.40\textwidth]{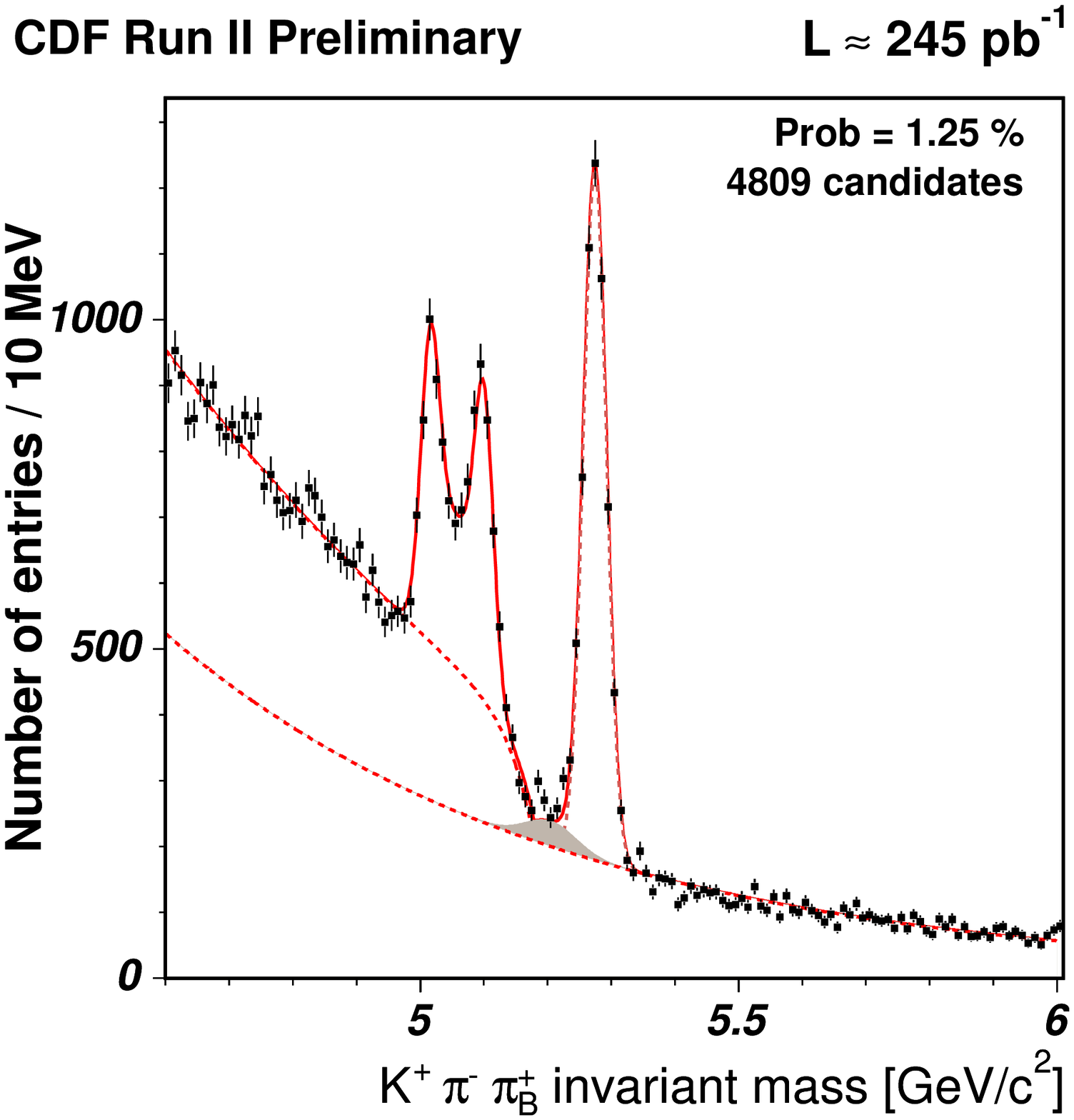} 
\hspace*{3mm}
\includegraphics*[width=0.40\textwidth]{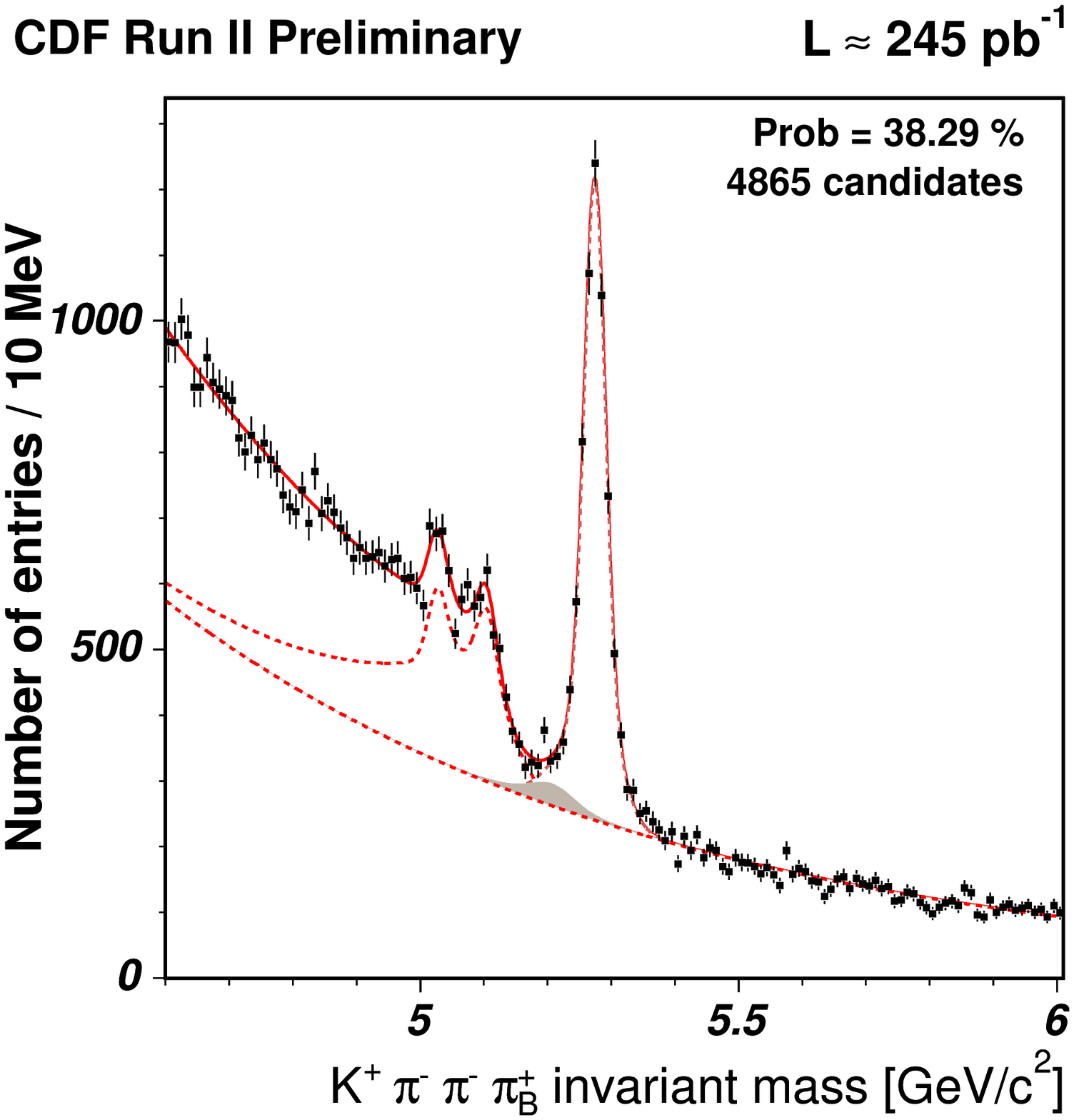}
\caption{%
Example of fully reconstructed $B$ meson signals 
among CDF data triggered with SVT.
Left: $B^- \rightarrow D^0 \pi^- \rightarrow (K^- \pi^+ ) \pi^-$.
Right:  $\bar B^0 \rightarrow D^+ \pi^- \rightarrow ( K^- \pi^+ \pi^+ ) \pi^-$.
}

\label{fig:svt-b-sig}
\end{center}
\end{figure}

\begin{figure}[thb]
\begin{center}
\includegraphics*[width=0.40\textwidth]{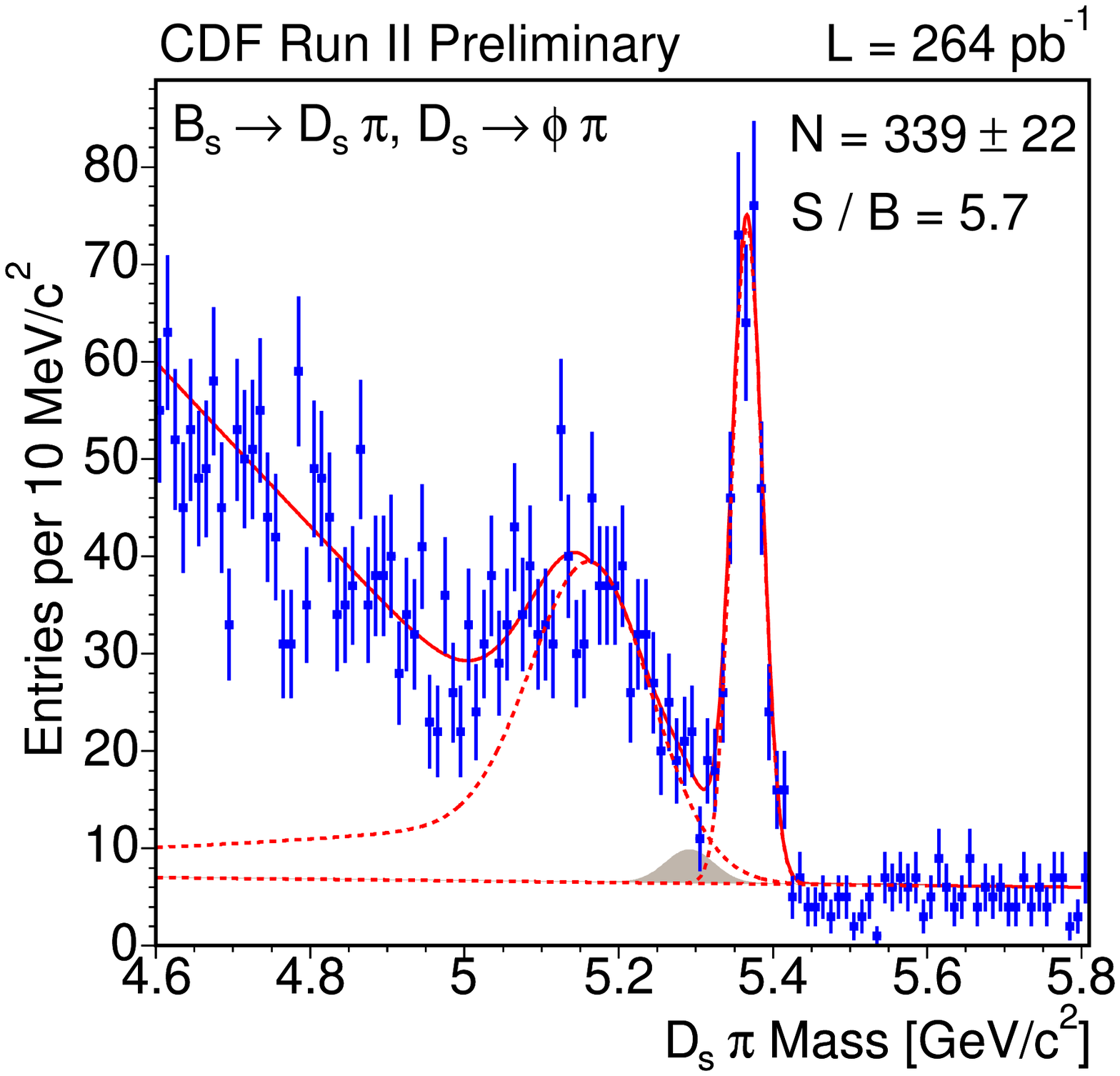} 
\hspace*{5mm}
\includegraphics*[width=0.36\textwidth]{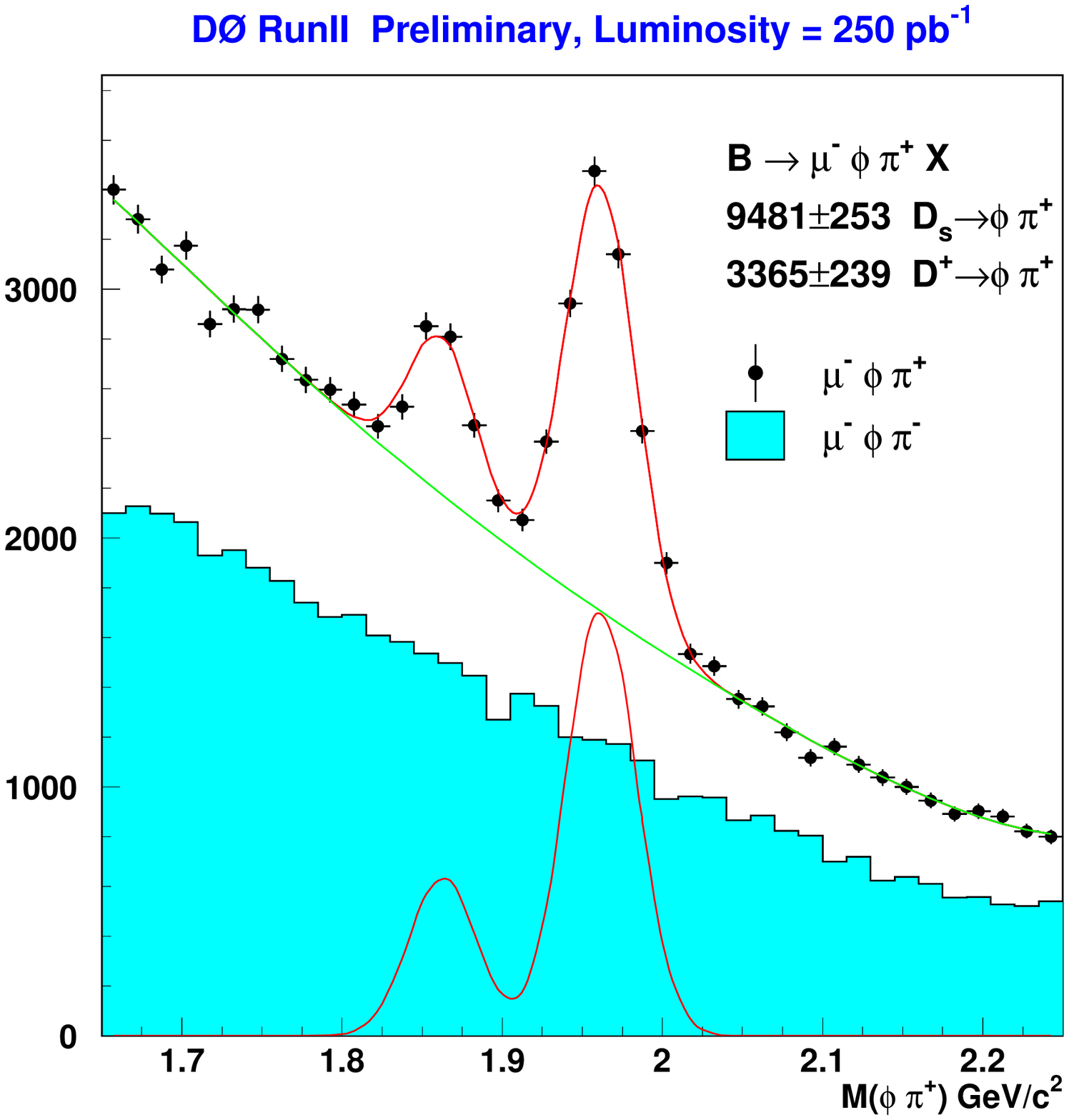}
\caption{%
Example of reconstructed $B^0_s$ meson signals. 
Left: $\bar B^0_s \rightarrow D^+_s \pi^- \rightarrow (\phi \pi^+ ) \pi^-$ (CDF).
Right:  $\bar B^0_s \rightarrow \mu^- \bar \nu  D^+_s X $ (D0).
}
\label{fig:bs-sig}
\end{center}
\end{figure}

\subsection{$B^0_s \bar B^0_s$ oscillations}

To study the consistency of the unitarity triangle 
of the KM matrix, 
the information on the lengths of the sides of the triangle
and their precise determination are crucial. 
In particular the determination of $ | V_{td} |$
from the oscillation frequency $\Delta m_d$ of $B^0_d \bar B^0_d$ mixing
currently
suffers from a relatively large theoretical uncertainty,
typically of order 20\%.
Unquenched lattice calculations of $B$ meson decay constants 
have become available in recent years, 
and they should help reduce the uncertainty.
From the experimental side, 
the theory uncertainty can in principle be reduced 
once we observe the $B^0_s \bar B^0_s$ oscillations
and use the ratio $\Delta m_s / \Delta m_d$
to extract $| V_{ts} / V_{td} |$.
For example, Ref.~\cite{lattice} calculates the ratio   $\xi$,  
of the $B^0_s$ and $B^0_d$ decay constants times the bag parameters, 
to be
$\xi  \equiv  ( f_{B_s} \sqrt{ B_{B_s} }  ) / 
              ( f_{B_d} \sqrt{ B_{B_d} }  ) 
 = 1.14 \, \pm \, 0.03 \, ^{ + \, 0.13} _{-\, 0.02}$, 
which involves a smaller uncertainty than 
when trying to extract $V_{td}$ using only $\Delta m_d$.
However, the expected very high value of $\Delta m_s$ 
poses a challenge
to experiments.

In order to measure particle-antiparticle oscillations
with a very high frequency, it is necessary, or at least desirable,
to have precise vertex determinations and 
a good proper time resolution. 
The former can be achieved by
having a good detector very close to the $B$ meson 
production point,
and the latter requires measuring $B$ meson momentum  on an event-by-event
basis. The latter can be achieved if decays are reconstructed fully. 
Such reconstruction of all-hadronic final states has become possible
in CDF Run-II, by using silicon detector information
at the second level of the trigger (SVT triggers). 
Figure~\ref{fig:svt-b-sig} shows an example of such signals.
They will be used as calibration modes, for understanding 
flavor tagging and proper time resolution.

The $B^0_s$ meson signals are also reconstructed by CDF with the SVT
trigger. Figure~\ref{fig:bs-sig} shows the signal, of about 340 events in
265~pb$^{-1}$ of data.
The study can in principle be made using 
the partially reconstructed semileptonic decay
$\bar B^0_s \rightarrow \ell^- \bar \nu D^+_s X$.
Figure~\ref{fig:bs-sig} (right) shows such a signal from the D0 experiment,
which takes advantage of a large acceptance of the muon detector.

Attempts will be made using these signals
to look for the oscillations
and possibly set a lower limit on $\Delta m_s$ toward winter 2005.

\begin{figure}[thb]
\begin{center}
\includegraphics*[width=0.40\textwidth]{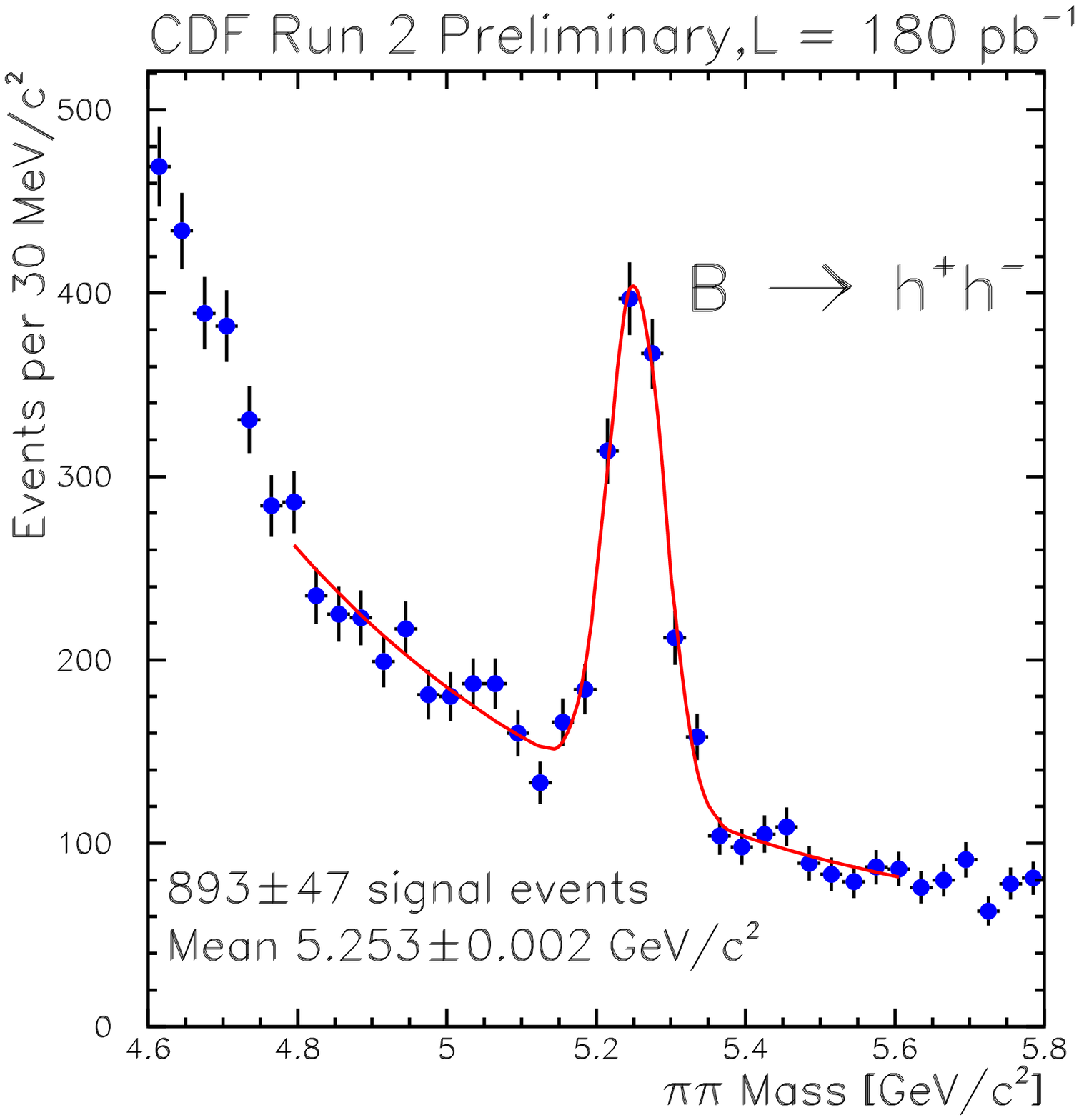} 
\hspace*{5mm}
\includegraphics*[width=0.40\textwidth]{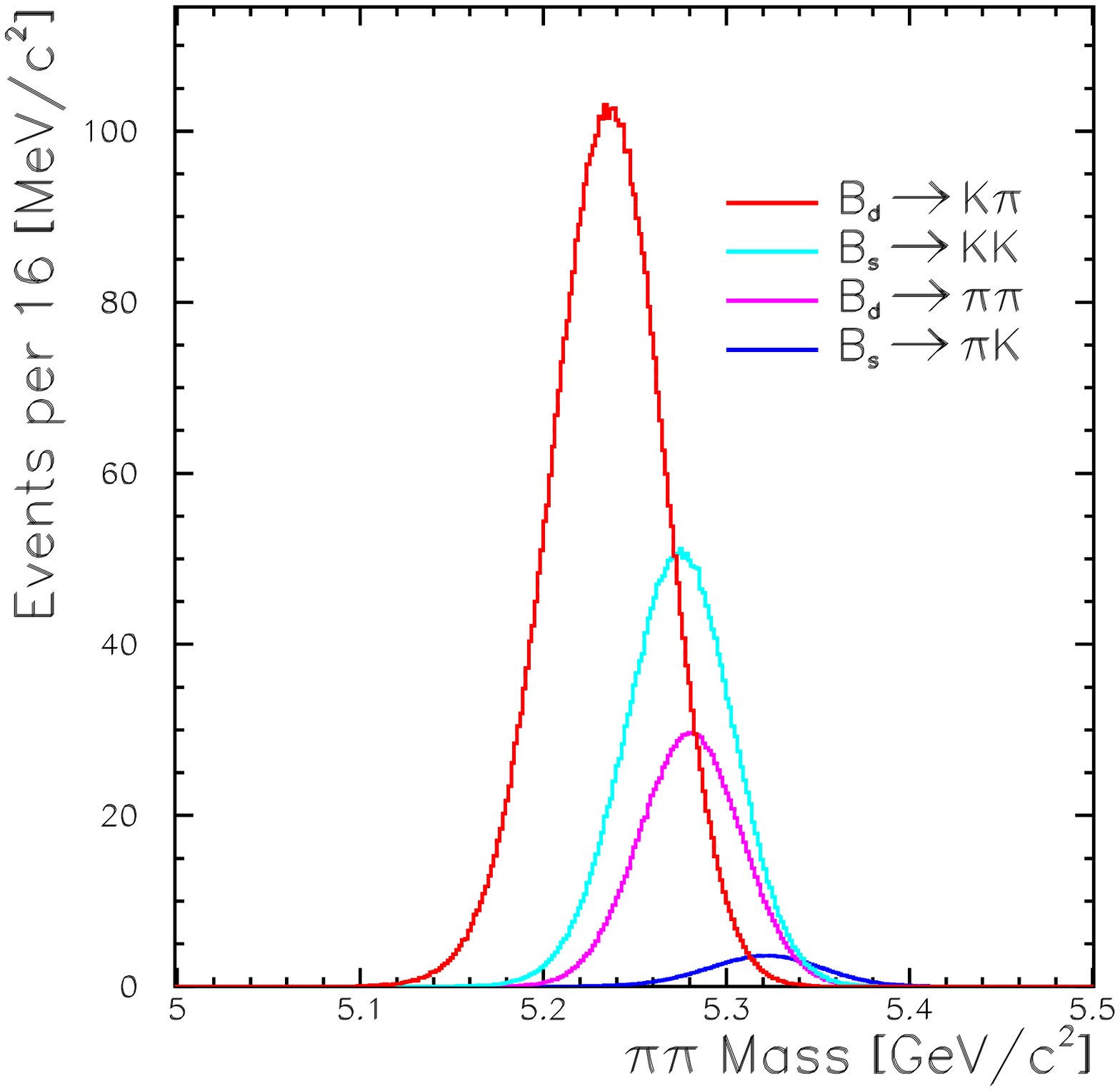}
\caption{%
Left: 
Invariant mass spectrum of two charged particles near the $B$ meson mass by CDF.
Pion masses are assigned to each particle. 
Right: Monte Carlo simulation of the mass spectra for four decay modes considered.
}
\label{fig:b-hh}
\end{center}
\end{figure}


\subsection{
Studies of $B^0_{d,s}  \rightarrow  P P $ decays }

The decay mode $B^0_d \rightarrow \pi^+ \pi^-$, 
if it proceeds only through a $b \rightarrow u$ ``tree"  transition,
receives the phase of $V_{ub}$ in the decay (which is angle $\gamma$)
and  the phase of $V_{td}$ in $B^0_d \bar B^0_d $ mixing (which is $\beta$).
Therefore, its CP asymmetry  should allow a determination
of $\sin 2 (\beta + \gamma)$, which should be identical 
to $\sin 2 \alpha$ if the triangle closes ($ \alpha + \beta + \gamma = \pi$).
However, the existence of the $b \rightarrow s $ ``penguin" amplitude 
complicates the matter, making it less straightforward
to extract
angle $\alpha$ from  experimentally observed CP asymmetry.
Various strategies have been proposed to solve this, but they are not necessarily 
easy experimentally. 

One proposed by Fleischer~\cite{fleischer} is unique in that
it measures      
CP asymmetries
in the $B^0_s \rightarrow K^+ K^-$ mode in  conjunction with   
the $B^0_d \rightarrow \pi^+ \pi^-$ mode, 
and extract angle $\gamma$ 
as well as tree and penguin decay amplitudes, 
taking advantage of the penguin pollution.
It is by no means easy experimentally, because it involves 
extraction of CP asymmetries modulated with $\Delta m_s$.

The first step toward those measurements will be 
to see the signals of
two-body decays of the $B$ mesons.
CDF uses a data sample triggered using SVT.
Figure~\ref{fig:b-hh} shows the invariant mass
spectrum of two-track pairs, where the pion mass is assumed
for both charged particles.
A clear signal of 900 events 
is observed near the $B$ meson mass.
The peak is actually expected to be a mixture of 
four decay modes, 
$B^0_d \rightarrow K^+ \pi^-$ and $ \rightarrow \pi^+ \pi^-$,
and 
$B^0_s \rightarrow K^+   K^-$ and $ \rightarrow  K^-  \pi^+$.
A Monte Carlo calculation of the mass spectra of these decay modes
is shown in Figure~\ref{fig:b-hh}~(right).
Specific ionization measurements ($dE/dx$) in the main tracking chamber 
are used to separate kaons and pions statistically 
and, together with the mass distributions, 
to estimate the mixture of the four decay modes.
The approximate yields are  
509 for $B^0_d \rightarrow   K^+ \pi^-$, 
134 for $B^0_d \rightarrow \pi^+ \pi^-$, 
and
232 for $B^0_s \rightarrow   K^+ K^-$.

This represents the  first observation of the decay $B^0_s \rightarrow K^+ K^-$. 
The ratio of the production fractions times the branching fractions is 
measured to be~\cite{b-hh}
\[
\frac { f(\bar b \rightarrow B^0_s ) \, \cdot
          {\cal B} ( B^0_s \rightarrow K^+ K^-   ) }
      { f(\bar b \rightarrow B^0_d ) \, \cdot \,
          {\cal B} ( B^0_d \rightarrow K^+ \pi^- ) }
= 
0.48  \pm 0.12 \pm 0.07 .
\]
Also the ratio of the branching fractions for the $B^0_d$ meson is extracted  to be
\[
\frac {    {\cal B} ( B^0_d \rightarrow \pi^+ \pi^- )   } 
        {    {\cal B} ( B^0_d \rightarrow K^+   \pi^- )   }  = 0.26 \pm 0.11 \pm 0.06  ,
\]
as well as direct CP asymmetry in the $B^0_d \rightarrow K^+ \pi^-$ decay 
\[
     {\cal A}_{\rm CP}( B^0 \rightarrow  K^+ \pi^- ) 
 \equiv  \frac {  \Gamma ( \bar B^0 \rightarrow K^- \pi^+ )  - \Gamma ( B^0 \rightarrow K^+ \pi^-  )   }
                   {  \Gamma ( \bar B^0 \rightarrow K^- \pi^+ )  + \Gamma ( B^0 \rightarrow K^+ \pi^-  )   }
= -0.04 \pm 0.08 \pm 0.01 .
\]
The latter is becoming competitive in precision   with the Belle and BaBar results.
In a longer term we hope to measure angle $\gamma$ with the Fleischer method.

\begin{figure}[thb]
\begin{center}
\includegraphics*[width=0.40\textwidth]{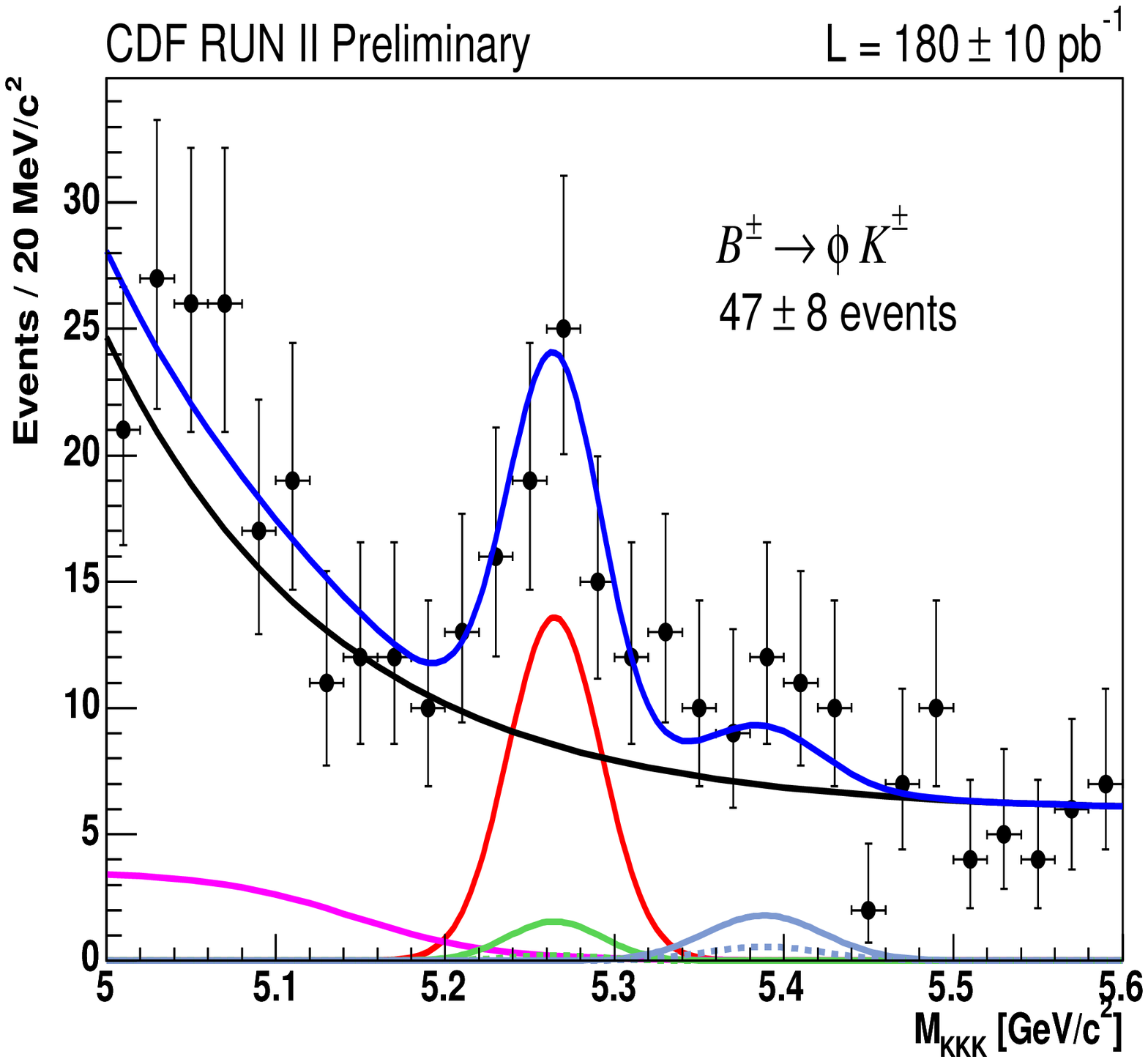} 
\hspace*{5mm}
\includegraphics*[width=0.40\textwidth]{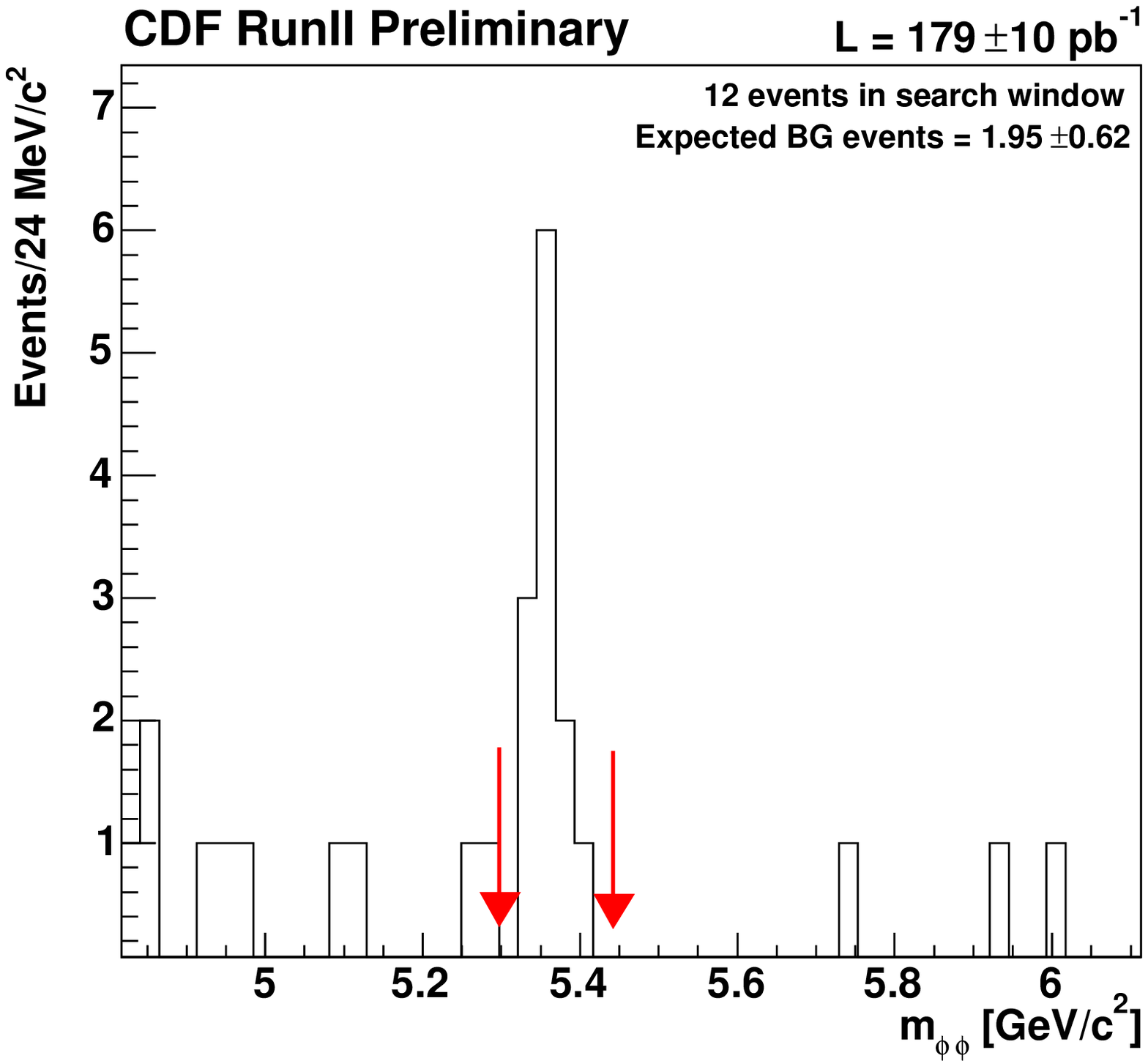}
\caption{%
Reconstructed CDF signals of $B^+ \rightarrow \phi K^+$ (left)
and $B^0_s \rightarrow \phi \phi$ (right).
}
\label{fig:phik}
\end{center}
\end{figure}

\subsection{Direct check with $B \rightarrow \phi K$ decays}

CDF reconstructs the  $B^+ \rightarrow \phi K^+$ decays
using the SVT trigger. 
Figure~\ref{fig:phik} shows the signal of about 50 events.
The ratio of branching fractions is measured to be
\[
\frac { {\cal B} ( B^+ \rightarrow \phi    K^+ ) }
      { {\cal B} ( B^+ \rightarrow J/ \psi K^+ ) }
  = ( 0.72 \pm 0.13 \pm 0.07 ) \times 10^{-2}  .
\]
The decay proceeds through the same quark level transition
$ b \rightarrow s \bar s s $ as the $B^0 \rightarrow \phi K^0_S$ mode.
Therefore if new physics phase exists it should show up in this mode
as well. Direct CP asymmetry is measured to be 
\[
  {\cal A}_{\rm CP} = 
 -0.07 \pm 0.17 \, ^{ +0.06} _ {-0.05}  , 
\]
which does not seem to show a large deviation from zero,
and 
again achieves a precision comparable to $B$ factory measurements.

Another $b \rightarrow s \bar s s $ transition mode has been seen at
CDF. It is the $B^0_s \rightarrow \phi \phi$ decay mode, whose
signal is shown in Figure~\ref{fig:phik} as well.

\section{Conclusion}
The Tevatron Run-II program is in progress since 2001.
Both CDF and D0 experiments have accumulated roughly five times more data than in Run I, 
with much improved detectors. 
Many physics results have been produced and more are expected in the near future.
They could be summarized as follows.
\begin{itemize}
\item Electroweak physics

Production of weak vector bosons has been measured 
at a new center of mass energy, and has provided
opportunities to study electroweak phenomena very precisely. 
Pairs of gauge bosons are now being produced in reasonably high statistics,
and interactions among gauge bosons will be studied in detail.
Measurement of the $W$ boson mass is
a high priority in the near term future.

\item Top quark physics 

Top quark pair production has been confirmed in Run II data. 
New precision in measurements of production cross sections and mass is expected and in some cases is already
achieved.
Many new different measurements will be performed, taking advantage of expected 20-fold increase
in the data sample and 30\% increase in the production cross section.
Combined with  $W$ boson mass measurements, top quark mass measurements
will provide indirect information on the Higgs boson mass.

\item Bottom quark physics

CDF has vastly improved its $B$ physics capability by introducing a displaced track trigger,
enabling to collect $B$ decays into final states consisting only of hadrons.
D0 is also commissioning a trigger under the same philosophy. \\
CDF and D0 could provide useful and unique measurements of $b \rightarrow s$
transitions, including $B^0_s \bar B^0_s$ oscillations, 
searches for $B^0_s \rightarrow \mu^+ \mu^-$ decays 
and non-zero width difference 
$\Delta \Gamma_s$, 
and
studies of decays $B^0_s \rightarrow J/\psi \phi$, $B^0_s \rightarrow K^+ K^-$,
and $B \rightarrow \phi K$.

\end{itemize}
In the near future, each of the CDF and D0 experiments is 
expected to collect about 
2~fb$^{-1}$ of integrated luminosity. 
We hope to see some exciting measurements come out of the data. 




\section{Acknowledgements}

I would like to thank the organizers of the Conference, 
in particular Professor Kaoru Hagiwara, Dr.~Nobuchika Okada 
and Dr.~Junichi Kanzaki
for their help, 
and  
also for 
patiently waiting for my writing this Proceedings contribution.
Many members of the CDF and D0 Collaborations 
helped me during the preparation of my talk and this manuscript.
They include Evelyn Thomson, 
Marco Verzocchi, 
Arnulf Quadt, 
Aurelio Juste, 
and Ralf Bernhard.



\end{document}